\documentclass[a4paper,11pt]{article}
\usepackage{siunitx}
\usepackage{jheppub}
\usepackage{array}
\usepackage{float}
\usepackage{booktabs}
\usepackage{makecell}
\usepackage{caption}
\usepackage{amsmath}
\usepackage{graphicx,tabularx}
\usepackage{gensymb}
\usepackage{url}
\usepackage{footmisc}
\usepackage{amsfonts}
\usepackage{cancel}
\usepackage{color}
\usepackage{multirow} 
\usepackage{amssymb}
\usepackage{pifont}
\usepackage{epstopdf}
\usepackage{slashed}
\usepackage{comment}
\usepackage{booktabs} 
\usepackage{subfigure}
\usepackage{mathrsfs}
\usepackage{subcaption}
\usepackage[toc,page]{appendix}
\usepackage{mathtools}
\usepackage{romannum}
\usepackage[normalem]{ulem}
\usepackage{bbold}
\usepackage{xcolor}
\usepackage{tikz-feynman}
\usepackage{tikz}
\usepackage{placeins}
\usepackage{hyperref}
\usepackage{orcidlink}
\allowdisplaybreaks

\usepackage{accents}

\def\beq{\begin{equation}}
\def\eeq{\end{equation}}
\title{
Probing Heavy Neutral Leptons through New Gauge Interactions at the Forward Physics Facilities
}

\author{Francis M. Burk$^1$\orcidlink{0009-0001-7398-8211},}
\affiliation{$^1$PITT PACC, Department of Physics and Astronomy, University of Pittsburgh, Pittsburgh, PA 15260, USA}
\emailAdd{frb25@pitt.edu}

\author{P. S. Bhupal Dev$^2$\orcidlink{0000-0003-4655-2866},}
\affiliation{$^2$Department of Physics and McDonnell Center for the Space Sciences, Washington University, St.~Louis, MO 63130, USA}
\emailAdd{bdev@wustl.edu}

\author{Bhaskar Dutta$^3$\orcidlink{0000-0002-0192-8885},}
\emailAdd{dutta@tamu.edu}
\affiliation{$^3$Mitchell Institute for Fundamental Physics and Astronomy, Department of Physics and Astronomy, Texas A\&M University, College Station, TX 77843, USA}

\author{Tao Han$^1$\orcidlink{0000-0002-5543-0716},}
\emailAdd{than@pitt.edu}

\author{Aparajitha Karthikeyan$^{3,4}$\orcidlink{0000-0002-6934-1239},}
\emailAdd{aparajitha\_96@tamu.edu}
\affiliation{$^4$Department of Physics and Astronomy, University of Iowa, Iowa City, IA 52242, USA}

\author{and Doojin Kim$^5$\orcidlink{0000-0002-4186-4265}}
\emailAdd{doojin.kim@usd.edu}
\affiliation{$^5$Department of Physics, University of South Dakota, Vermillion, SD 57069, USA}

\preprint{
\begin{flushright}
PITT-PACC-2617\\
FERMILAB-PUB-26-0766\\MI-HET-895
\end{flushright}
}

\abstract{
We investigate the prospects for improving the heavy neutral lepton (HNL) searches at the far-forward LHC experiments FASER and FASER$\nu$, as well as their proposed HL-LHC successors, FASER2 and FASER$\nu$2, using new HNL production mechanisms. 
We consider scenarios in which HNLs couple directly to a new $U(1)$ gauge boson and decay into visible Standard Model final states through mixing with active neutrinos. Focusing on the $U(1)_{B-L}$, $U(1)_{B-3L_\tau}$, and $U(1)_B$ gauge extensions as benchmark models, we study HNL production through proton bremsstrahlung and Drell--Yan processes, which probe complementary mediator-mass regimes. We analyze several visible decay channels, including charged leptons, neutral mesons, and other hadrons, and evaluate the projected sensitivities for different neutrino-flavor mixing patterns. We find that these new production mechanisms that differ qualitatively from the conventional picture in which both production and decay are controlled by active-sterile mixing enable the far-forward LHC program to probe previously unexplored regions of HNL parameter space, with the first-ever demonstration of FASER$\nu$ having sensitivity to HNLs and FASER2 and FASER$\nu$2 offering substantially enhanced sensitivity during the HL-LHC era, even reaching the canonical seesaw prediction for all three lepton flavors. 
Our results demonstrate that new gauge interactions can substantially enhance HNL production independently of the active-sterile mixing that governs their visible decays, opening complementary opportunities to explore HNLs at the forward physics facilities.  
}

\begin{document}

\titlepage
\maketitle

\section{Introduction}
Heavy neutral leptons (HNLs) provide a well-motivated extension of the Standard Model (SM), with implications for neutrino mass generation~\cite{Abdullahi:2022jlv} and possible connection to dark sectors~\cite{Batell:2022xau}. As SM singlets, HNLs can mix with active neutrinos, allowing them to decay into visible SM final states. Their phenomenology can be further enriched in the presence of a new neutral gauge boson ($Z'$) that couples directly to HNLs and SM fermions. Such mediators arise naturally in a variety of extensions of the SM gauge group, including $U(1)_B$, $U(1)_{B-L}$, and $U(1)_{B-3L_i}$.  In these scenarios, HNL production and decay need not be governed by the same interaction: a new gauge interaction can enhance production, while active-sterile mixing controls the subsequent decay into SM particles.

In this work, we focus on the GeV-scale HNL parameter space, which can be explored through a variety of fixed-target, beam-dump, and collider experiments, including 
the ongoing SBND~\cite{MicroBooNE:2015bmn}, ICARUS~\cite{ICARUS:2023gpo}, DarkQuest~\cite{Apyan:2022tsd}, and FASER~\cite{FASER:2022hcn} experiments, as well as future facilities such as DUNE~\cite{DUNE:2021tad} and SHiP ~\cite{SHiP:2021nfo}.  Searches at these experiments are complementary to those at the LHC, which primarily probes new physics at mass scales of $\mathcal{O}(100~\mathrm{GeV})$ to a few TeV.

Recently, we demonstrated the potential for exploring HNLs in $U(1)$ extensions at SBND, ICARUS, DarkQuest, DUNE, and SHiP by utilizing neutral meson decays, proton bremsstrahlung, and Drell-Yan scattering processes~\cite{Capozzi:2024pmh, Burk:2026mvk}. In these experiments, proton beams with energies ranging from 8 GeV to 400 GeV impinge on a fixed target, providing a unique opportunity to produce HNLs beyond the conventional production mechanisms based primarily on charged-meson decays. Their subsequent decays inside the detector into final states containing electrons, muons, neutral pions, and other particles, offer distinctive experimental signatures. We showed that these additional production mechanisms significantly extend the regions of parameter space that can be probed compared to conventional HNL searches. 

Complementarily, the far-forward region of the LHC offers an opportunity to probe HNLs produced in high-energy proton--proton collisions, with their long lifetimes allowing them to travel substantial distances before decaying. In this work, we investigate the sensitivity of the ongoing FASER and FASER$\nu$ experiments~\cite{FASER:2018bac, FASER:2020gpr}, as well as their proposed high-luminosity (HL) LHC successors, FASER2 and FASER$\nu2$~\cite{Feng:2022inv}, to HNLs produced through a $Z'$ mediator. The FASER setup allows us to exploit not only the production of HNLs in proton--proton collisions at the ATLAS interaction point (IP), but also their production from secondary neutrons interacting with the TAN and TAXN absorbers downstream of the ATLAS IP~\cite{Dev:2023zts}. We utilize the FASER detector together with the FASER$\nu$ and FASER$\nu2$ detectors to search for HNL decays into visible final states containing electrons, muons, tau leptons, neutral mesons, and other particles.
We include two complementary production mechanisms: proton bremsstrahlung, relevant for relatively light mediators, and Drell-Yan production, which becomes important for heavier mediators.

For the mediator $Z'$ interaction, we consider three representative gauge symmetries, $U(1)_{B-L}$, $U(1)_{B-3L_\tau}$, and $U(1)_B$, and examine both flavor-democratic and $\tau$-dominated HNL mixing scenarios. Our results show that mediator-assisted production enables the far-forward LHC experiments to probe regions of HNL parameter space beyond existing constraints/projections~\cite{Kling:2018wct,Feng:2024zfe}, with FASER2 and FASER$\nu$2 offering particularly promising reach during the HL-LHC era. We find that production at the ATLAS IP dominates over the secondary production at the TAN/TAXN absorbers in the scenarios considered. More broadly, this study demonstrates how additional gauge interactions can open new opportunities for HNL searches by enhancing their production independently of the small mixing that governs their visible decays.

The rest of the paper is organized as follows. Section~\ref{sec:model} introduces the theoretical framework featuring a new vector mediator and HNL. Section~\ref{sec:Exp_Settings} summarizes the relevant experimental configurations. Section~\ref{sec:sigcalc} describes the production mechanisms, signal selection, and projected sensitivities, including the contribution from secondary interactions at the TAN/TAXN absorbers. Section~\ref{sec:conclusion} presents our conclusions. Appendices~\ref{app:brem} and~\ref{App:Secondary_Fluxes} give some details of the HNL production from bremsstrahlung and from primary neutrons and secondary hadrons at TAN/TAXN, respectively. 

\section{Theoretical Model Setting}
\label{sec:model}
\subsection{Gauge Interactions and Benchmark Models}
\label{sec:gauge}
We describe the interactions relevant to our analysis using an effective theory containing a massive neutral vector boson collectively called $Z'$ and a Majorana heavy neutral lepton denoted by $N$. The new boson mediates interactions between the HNLs and the SM fermions, while mixing between the HNLs and the active neutrinos allows $N$ to decay into visible SM final states.  
The effective Lagrangian is expressed as
\begin{equation}
    \mathcal{L}_X \supset -\frac{1}{4}F'_{\mu\nu}F'^{\mu\nu} + \frac{1}{2}m_{Z'}^2Z_{\mu}'Z'^{\mu} +g_{X}Z_{\mu}'\Bigg(\sum_f\bar{f}\gamma^{\mu}(g_{V,f} + g_{A,f}\gamma_5)f +\bar{N}\gamma^{\mu} (g_{V,N} + g_{A,N}\gamma_5) N\Bigg),
    \label{eq:lag}
\end{equation}
where $F'_{\mu\nu}=\partial_\mu Z'_\nu-\partial_\nu Z'_\mu$ determines the kinetic term, $m_{Z'}$ denotes the mediator mass, $g_X$ is the overall gauge coupling, and $f=q,l$ is summed over all SM quarks and leptons. The coefficients $g_{V,f}$ and $g_{A,f}$ specify the vector and axial-vector charges of each fermion under the new interaction. For a Majorana HNL, the diagonal vector current vanishes, so its coupling to $Z'$ is purely axial-vector in this parameterization.

We consider three representative gauge assignments: $U(1)_{B-L}$, $U(1)_{B-3L_\tau}$, and $U(1)_B$. They provide distinct patterns of couplings to quarks and leptons and consequently different constraints on the mediator parameters. The charge assignments used in our calculations are summarized in Table~\ref{ModelCouplings}.
\begin{table}[tb]
\centering
\setlength{\tabcolsep}{1.5pt}  
\renewcommand{\arraystretch}{1.3} 
\footnotesize
\begin{tabularx}{0.5\linewidth}{X || c c ||c c || c c }
\toprule
  ~~~ & $g_{V,q}$ & ~~~$g_{A,q}$ & $g_{V,l}$ & ~~~$g_{A,l}$& $g_{V,N}$ & ~~~$g_{A,N}$ \\
\midrule
$U(1)_{B-L}$ & 1/3  & 0 & $-1$ & 0 & 0 & $-1$ \\
$U(1)_{B-3L_{\tau}}$  & 1/3  & 0 & $-3$ & 0 & 0 & $-1$ \\
$U(1)_{B}$  & 1/3  & 0 & 0 & 0 & 0 & $-1$ \\ 
\bottomrule
\end{tabularx}
\caption{Fermion couplings for the benchmark gauge assignments. The lepton coupling in the $U(1)_{B-3L\tau}$ row applies to the $\tau$ flavor only; the other SM lepton flavors are uncharged under this symmetry. The $U(1)_B$ benchmark has no direct $Z'$ coupling to SM leptons. The vector coupling of HNL is zero because it is assumed to be Majorana.}
\label{ModelCouplings}
\end{table}
These benchmarks span qualitatively different experimental signatures. In the $B-L$ scenario, the mediator couples to both quarks and leptons. The $B-3L_\tau$ realization singles out the third lepton generation, while the baryonic $U(1)_B$ model couples the mediator to quarks but not directly to SM leptons. The latter therefore provides a useful benchmark in which constraints relying on visible dilepton decays are absent or substantially weakened. Note that the vector coupling of HNL ($g_{V,N}$) is zero because it is assumed to be Majorana. Consistent gauge theories may require additional fermions or other contributions to cancel gauge anomalies; we use the benchmark interactions above as an effective description of the phenomenology of interest and ignore the other possible heavier degrees of freedom in the current consideration.

Existing searches constrain the allowed mediator masses and couplings in a model-dependent fashion. For $U(1)_{B-L}$, relevant bounds include searches for light vector states in electron--positron collisions and rare decays. In the $U(1)_{B-3L_\tau}$ case, constraints from non-standard neutrino interactions are particularly relevant. The baryonic scenario is constrained by a combination of rare meson decays, quarkonium decays, fixed-target measurements, and searches sensitive to additional states required by the ultraviolet completion. We impose these constraints when selecting benchmark couplings for the sensitivity analysis in Section~\ref{sec:Sensitivity}. Further details of the underlying models and existing bounds can be found in Refs.~\cite{ Ilten:2018crw, Bauer:2018onh, Burk:2026mvk}.

\subsection{Active-sterile Mixing and Benchmark Choices}
\label{sec:HNLmixing}
In addition to the gauge interaction in Eq.~\eqref{eq:lag}, the HNL is allowed to mix with the SM active neutrinos. In the mass basis, the flavor eigenstates can be expressed as
\begin{equation}
    \nu_\alpha = \sum_{i=1}^3 U_{\alpha i}\nu_i+U_{\alpha N}N \, , \qquad
\alpha=e,\mu,\tau.
\end{equation}
where $\nu_i$ are the three light-neutrino mass eigenstates. We write $U_\alpha\equiv U_{\alpha N}$ and define the total mixing strength as
\begin{align}
|U|^2\equiv |U_e|^2+|U_\mu|^2+|U_\tau|^2 \, .
\label{eq:Utot}
\end{align}
We treat $m_N$ and the mixing parameters as phenomenological inputs and do not specify a particular ultraviolet completion of the neutrino-mass sector. Although realistic neutrino-mass models generally contain at least two HNLs, a single accessible state provides an adequate benchmark when additional states are kinematically inaccessible or sufficiently close in mass that their individual contributions need not be resolved.

We examine two limiting patterns of flavor mixing:
\begin{align}
\text{Flavor-democratic (111):}& \quad
(|U_e|^2,|U_\mu|^2,|U_\tau|^2)
=|U|^2\left(\frac13,\frac13,\frac13\right), \label{eq:111} \\
\text{Tau-dominated (001):}&\quad
(|U_e|^2,|U_\mu|^2,|U_\tau|^2)
=|U|^2(0,0,1). \label{eq:001}
\end{align}
These benchmarks capture complementary possibilities for HNL decays and detection efficiencies. Equal mixing case is allowed by current global fits to neutrino oscillation data in the inverted-ordering scenario~\cite{Chrzaszcz:2019inj, Drewes:2022akb}. On the other hand, the $\tau$-dominated mixing is motivated by neutrino-mass scenarios that permit a large $\tau$-flavor component that is currently allowed at the $3\sigma$ level by the global oscillation data~\cite{Chrzaszcz:2019inj, Drewes:2022akb} and illustrates the reach in a sector for which experimental constraints are comparatively weak.

For comparison with existing HNL limits (see e.g., Refs.~\cite{Atre:2009rg, Deppisch:2015qwa, deGouvea:2015euy, Cai:2017mow, Bolton:2019pcu}), we use the constraints appropriate to each mixing benchmark. For the $(001)$ benchmark, the current limits on the HNL mixing versus its mass are taken from Refs.~\cite{bolton, hostert}. For GeV-scale HNLs, the relevant limits on $U_{\tau}$ come from CHARM~\cite{Orloff:2002de, Boiarska:2021yho}, BEBC~\cite{WA66:1985mfx, Barouki:2022bkt},  ArgoNeuT~\cite{ArgoNeuT:2021clc}, BaBar~\cite{BaBar:2022cqj}, and DELPHI~\cite{DELPHI:1996qcc}.  For the $(111)$ benchmark, we combine flavor-dependent limits using the corresponding flavor fractions, while retaining published constraints from flavor-inclusive searches. Bounds from flavor-blind searches, such as DELPHI~\cite{DELPHI:1996qcc}, are applied without rescaling. For HNLs with masses $m_N\lesssim 2 \text{ GeV}$, we adopt the limits obtained in Ref.~\cite{Feng:2024zfe} to recast the dominant constraints from T2K~\cite{T2K:2019jwa, Arguelles:2021dqn} and CHARM~\cite{Boiarska:2021yho}. The remaining dominant constraints for this benchmark arise from searches by ATLAS and CMS. The recent CMS~\cite{CMS:2024ita} and ATLAS~\cite{ATLAS:2022atq,ATLAS:2025uah} searches provide limits directly for the $(111)$ benchmark, so no recast is necessary.  On the other hand, some ATLAS and CMS searches assume one flavor dominance. To accommodate the new benchmarks, one could conduct a full reanalysis of the data from experiments, as done in e.g.,  Ref.~\cite{Tastet:2021vwp}. A simpler alternative is to recast the published ATLAS and CMS limits using approximate scalings of the signal rate with the mixing parameters, which is sufficient for our purposes. 
Following Ref.~\cite{Abada:2022wvh}, we write $U_{\alpha}^2 = \varepsilon_{\alpha}^2 U^2$, with $\sum_{\alpha = e, \mu, \tau}\varepsilon_{\alpha}^2 = 1$, so that $U^2$ is the total mixing and $\varepsilon_\alpha^2$ is the fraction carried by flavor $\alpha$. LHC searches for heavy Majorana neutrinos with $m_N \geq 1~\text{GeV}$ typically target the final states given below. Assuming prompt decays and neglecting lepton masses, their rates scale roughly as:
\begin{align}
\sigma(pp \to \ell_{\alpha}^{\pm}\ell_{\alpha}^{\pm} + {\rm jets}) &\propto U^2|\varepsilon_{\alpha}|^4, \label{resc_eqn1}\\
\sigma(pp \to \ell_{\alpha}^{\pm}\ell_{\beta}^{\pm} + {\rm jets}) &\propto U^2|\varepsilon_{\alpha}|^2|\varepsilon_{\beta}|^2, \label{resc_eqn2}\\
\sigma(pp \to \ell_{\alpha}^{+}\ell_{\alpha}^{+}\ell_{\alpha}^- + \slashed{E}_{T}) &\propto U^2|\varepsilon_{\alpha}|^4, \label{resc_eqn3}\\
\sigma(pp \to \ell_{\alpha}^+\ell^{+}_{\alpha}\ell^-_{\beta} + \slashed{E}_T) &\propto U^2|\varepsilon_{\alpha}|^2\left(|\varepsilon_{\alpha}|^2 + |\varepsilon_{\beta}|^2\right), \label{resc_eqn4}\\
\sigma(pp \to \ell_{\alpha}^+\ell^{-}_{\alpha}\ell^+_{\beta} + \slashed{E}_T) &\propto U^2|\varepsilon_{\alpha}|^2\left(|\varepsilon_{\alpha}|^2 + 3|\varepsilon_{\beta}|^2\right). \label{resc_eqn5}
\end{align}
We rescale the ATLAS limit of Ref.~\cite{ATLAS:2024fcs} using Eq.~\eqref{resc_eqn1}, and those of Ref.~\cite{ATLAS:2025lva} using Eq.~\eqref{resc_eqn4}. The relevant CMS searches~\cite{CMS:2018iaf, CMS:2024xdq} quote limits from a combination of trilepton final states. For these, we rescale using Eq.~\eqref{resc_eqn5}. In each case, we rescale the most constraining of the published single-flavor limits. For the CMS displaced-vertex searches of Refs.~\cite{CMS:2022fut, CMS:2024ake}, we show the published limit without rescaling. The limit's upper boundary is set by the requirement that $N$ decays sufficiently far from the primary vertex, and is therefore controlled by the $N$ lifetime, which depends to a good approximation only on the total mixing $U^2$.


\section{Forward Physics Experimental Settings}
\label{sec:Exp_Settings}
As shown below, far-forward experiments provide sensitivity to HNLs produced through both bremsstrahlung and Drell-Yan processes. We consider four detector configurations located along the beam collision axis downstream of the ATLAS IP: FASER~\cite{FASER:2018bac,FASER:2022hcn} and FASER$\nu$~\cite{FASER:2020gpr,FASER:2026jhh}, which are currently operating in the TI12 service tunnel of the LHC experiment, and their proposed successors FASER2 and FASER$\nu$2 ~\cite{Feng:2022inv,FPF:2025bor}, which will be housed in the Forward Physics Facility (FPF) during the HL-LHC era. 
Since the FPF detector designs are still being optimized, we adopt the baseline specifications of the current conceptual design. The detector dimensions and integrated luminosities relevant to our geometric and decay-acceptance calculations are summarized below and in Table~\ref{tab:expsummary}.

\begin{table}[ht!]
    \centering
    \begin{tabular}{c | c c c c}
    \toprule
     Experiment & Baseline & Transverse size & Fiducial decay length & Luminosity  \\
     \midrule
    FASER~\cite{FASER:2022hcn, kose} & 480~m & $R=10$~cm & 1.5~m & 343~fb$^{-1}$  \\
    FASER$\nu$~\cite{FASER:2026jhh, kose} & $\sim480$~m & $25\times 30$~cm$^2$ & 1.0~m & 240~fb$^{-1}$ \\
    FASER2~\cite{FPF:2025bor, vbbq-fb1t,MammenAbraham:2025gai} & $\sim650$~m & $2.6\times 1$~m$^2$ & 10~m & 3~ab$^{-1}$ \\
    FASER$\nu$2~\cite{FPFWorkingGroups:2025rsc, Adhikary:2024nlv} & $\sim640$~m & $40\times 40$~cm$^2$ & 6.6~m & 3~ab$^{-1}$ \\
     \bottomrule
    \end{tabular}
    \caption{Summary of key parameters of the four forward physics experiments considered in this study. For FASER and FASER$\nu$, the quoted integrated luminosities correspond to the accumulated datasets collected for these experiments to date, which exceed their original Run-3 target exposures. For FASER2 and FASER$\nu$2, we use the projected HL-LHC luminosity.}
    \label{tab:expsummary}
\end{table}

\begin{itemize}
    \item[$\bullet$] FASER~\cite{FASER:2022hcn}: The FASER experiment is a compact forward experiment, which was originally designed to search for light, weakly-interacting particles throughout Run 3 of the LHC from 2022-26. 
    Located $480$ m downstream of the ATLAS IP, the experiment is sensitive to particles which decay within a cylindrical volume of radius $R = 10$ cm and length $L = 1.5$ m. In addition, there is a planned photon preshower upgrade slated to be placed roughly $2.7 \text{ m}$ downstream of the decay volume. Events are resolved by an electromagnetic calorimeter (ECAL). The experimental design features a neutral particle absorber TAN located $L_0 = 130$ m downstream of the ATLAS IP. Neutrons impinging on the absorber may scatter with the iron in the target, producing additional secondary fluxes~\cite{Dev:2023zts}.

    \item[$\bullet$] FASER$\nu$~\cite{FASER:2026jhh}: FASER$\nu$ is a small emulsion detector placed directly in front of the FASER apparatus designed to operate during the same data-taking period as FASER. The detector is composed of 730 emulsion layers interleaved with tungsten plates. This yields an effective decay volume of $25 \text{ cm} \times 30 \text{ cm} \times1.0 \text{ m}$.

    \item[$\bullet$] FASER2~\cite{FPF:2025bor, vbbq-fb1t,MammenAbraham:2025gai}: The FASER2 experiment is a larger successor of FASER designed to operate during the HL-LHC run, collecting an expected integrated luminosity of $\mathcal{L} = 3000 \text{ fb}^{-1}$ of data. The decay volume is proposed to be located roughly $650$ m downstream of the ATLAS IP, within the dimensions of a $2.6 \times 1 \times 10 \text{ m}^3$ cuboid. 
    Similar to FASER, the experiment will include a photon preshower station at roughly $10$ meters after the decay volume, and the electromagnetic energy of events will be measured by an ECAL.
    The experimental design features a neutral particle absorber TAXN located $L_0 = 130$ m downstream of the ATLAS IP. Neutrons impinging on the absorber may interact with the iron nuclei in the target, producing additional secondary fluxes~\cite{Dev:2023zts}.

    \item[$\bullet$] FASER$\nu$2~\cite{FPFWorkingGroups:2025rsc, Adhikary:2024nlv}: FASER$\nu$2 is designed to be a much larger successor to FASER$\nu$ with proposed data taking occurring during the same time as FASER2's operation, and it will be located roughly $640 \text{ m}$ downstream from the ATLAS IP. The detector is envisioned 
    as a 20-ton tungsten-emulsion detector comprising approximately 3300 emulsion layers interleaved with 2-mm-thick tungsten plates. The tungsten target has approximate dimensions $40 \text{ cm} \times 40 \text{ cm} \times 6.6  \text{ m}$.
    
\end{itemize}

\section{Signal Calculation, Event Selection, and Sensitivity Reach}
\label{sec:sigcalc}
We study the sensitivity of the FASER experiments and their FPF successors to HNLs produced from the decay of a massive $Z'$. The signal may be produced at the ATLAS IP 
and downstream in the neutral-particle absorber (TAN in Run 3 and TAXN at the HL-LHC) as secondary fixed target collisions. 
We find that the signal production at ATLAS is significantly larger than the secondary fixed target production. We will thus describe the signal calculation at the IP in this section, and leave the subdominant production at the fixed target to Appendix~\ref{App:Secondary_Fluxes}.

At the ATLAS IP, the number of events produced can be simply expressed as
\begin{equation}
N_{\text{events}} = \sigma( X)\times\mathcal{L}
\label{eq:Nev}
\end{equation}
where $\sigma(X)$ is the production cross section of the BSM process $X$ of our interest and $\mathcal{L}$ is the integrated luminosity. To estimate the number of events that can be detected in a detector, one includes the branching ratio to the observable SM final state BR($X\to \text{final states})$ with detection efficiency $\text{eff}_f$ to find 
\begin{equation}
N_{\text{Detected}} = N_{\text{events}}\times \text{BR}(X\to \text{final states})\times \text{eff}_f \ .
\label{eq:ndet}
\end{equation}
For $N_{\text{MC}}$ Monte-Carlo-generated events, the detection efficiency is calculated by~\cite{Berlin:2018pwi}
\begin{equation}
\text{eff}_f = m\Gamma\int_{z_{\text{min}}}^{z_{\text{max}}}dz\sum_{\text{events}~\in~ \text{geom.}}\frac{e^{-z(m/p_z)\Gamma}}{N_{\text{MC}}\ p_z},
\label{eq:eff}
\end{equation}
where $z_{\min}$ and $z_{\max}$ define the length scale of the ``effective'' decay volume along the beam axis, with the corresponding component of the decaying particle's momentum $p_z$. Here $\Gamma$ is the total decay rate.

The energy scale and baseline of the FASER experiments make them sensitive to light dark-sector particles~\cite{FPF:2025bor} with masses ranging from $\mathcal{O}(1~{\rm MeV})$ to $\mathcal{O}(10~{\rm  GeV})$. However, lighter HNLs typically have longer lifetimes and decay lengths, rendering FASER less sensitive to $m_N \lesssim 100~\rm{MeV}$, for which the decay lengths can exceed the detector baseline. Since we focus on HNLs produced through the decays of mediators, we restrict our analysis to mediator masses heavier than $m_{Z'} \gtrsim \mathcal{O}(300~\rm{MeV})$, or correspondingly, $m_{N} > \mathcal{O}(100~\rm{MeV})$, where the two main production mechanisms are bremsstrahlung and Drell-Yan production. We describe them and our treatment as follows.\footnote{While exotic two- and three-body meson decays are also a relevant production mechanism for $Z'$~\cite{Ilten:2018crw, FASER:2023tle}, they are subdominant in the mass regime of interest.}

\subsection{Bremsstrahlung Production}
\label{sec:brem}

Gauge boson mediators at or below the proton mass scale ($m_{Z'} \lesssim 2$~GeV) are dominantly produced from bremsstrahlung, a long-considered source of high-energy, forward-directed mediators with a sizable cross section. As with lepton bremsstrahlung, the Fermi-Weiszacker-Williams approximation~\cite{Blumlein:2013cua, Gorbunov:2014wqa} has been widely utilized to estimate sensitivities and signals at various beam-dump experiments~\cite{deNiverville:2016rqh, Berryman:2019dme}. However, the non-perturbative nature of the strong interaction requires accounting for effects such as Pomeron exchange~\cite{Foroughi-Abari:2021zbm}, which reveal that the dominant contribution comes from non-single diffractive (NSD) scattering, with gauge bosons emitted only via quasi-real initial-state radiation (QR-ISR). This framework was later extended to treat soft divergences~\cite{Foroughi-Abari:2024xlj} and to provide simplified form factors for a broad class of BSM $U(1)$ models~\cite{Kling:2025udr}. In this analysis, we adopt the ISR bremsstrahlung treatment of Ref.~\cite{Kling:2025udr}. 
For an incoming nucleon (with four-momentum $p_N$) that impinges on a target ($p_t$) to give rise to NSD states $f$ along with a vector meson $V$, the differential cross section for ISR bremsstrahlung can be written as the NSD cross section times a splitting function $\omega(z, p_T^2)$: 
\begin{equation}
    d\sigma^{p_N p_t \to V f} = \omega_N(z, p_T^2) dz dp_T^2\times \sigma_{pp}^{\rm{NSD}}(s') , 
    \label{eq:bremcs}
\end{equation}
where $z = k_z/(p_N)_z$ is the fraction of longitudinal momentum that is transferred from the incoming nucleon to the outgoing $V$ that has a momentum $k^\mu$, $p_T^2 = k_x^2 + k_y^2$ is the transverse momentum-squared of the outgoing $V$, and $s' = (p_N-k+p_t)^2$ is the underlying center-of-mass (COM)  energy of the scattering process when the intermediate proton undergoes NSD scattering. The splitting function also includes the Breit-Wigner Dirac and Pauli form factors as well as a timelike form factor that ensures that the intermediate nucleon is close to being on-shell. We defer the reader to Appendix~\ref{app:brem} for more details on the bremsstrahlung formulation and kinematic conditions. Using the differential cross section in Eq.~\eqref{eq:bremcs}, we then calculate the events from the ATLAS IP using Eqs.~\eqref{eq:Nev} and~\eqref{eq:ndet}.

\subsection{Drell-Yan Production}
\label{sec:DY}

For heavier $Z'$ with $m_{Z'} \gtrsim 2$ GeV, 
the large momentum transfer opens the new production process via the quark-antiquark ($q\bar q$) annihilation, namely the partonic ``Drell-Yan'' process for a resonant production of $Z'$. 
The incoming $q$ and $\bar{q}$ carry some momentum fractions $x_1$ and $x_2$ of their parent protons. The squared-energy in the partonic COM frame is $\hat{s} \equiv \tau s = x_1x_2s$. 
The cross section of $X$ final state production from incoming hadrons $h$ on a proton $p$ in terms of the partonic luminosity $\mathcal{L}_{i,j}$ can be written as 
\begin{align}
\sigma(hp\to X) &= \int_{\tau_0}^1d\tau\sum_{q,\bar{q}}\frac{d\mathcal{L}_{q,\bar{q}}}{d\tau}\,\hat{\sigma}(q\bar{q}\to X), \quad \tau_0= {m^2_X\over s}, \\
 \frac{d\mathcal{L}_{i,j}}{d\tau} &= \frac{1}{1 + \delta_{ij}}\int_{\tau}^1\frac{d\xi}{\xi}\left[f_{i/h}\big(\xi,Q^2\big)f_{j/p}\left(\frac{\tau}{\xi},Q^2\right) + (i\leftrightarrow j)\right] ,
\end{align}
where $Q^2$ is the QCD factorization scale and
$f_{i/h}, f_{j/p}$ are the parton distribution functions associated with the $i$th and $j$th partons of hadrons $h$ and $p$, respectively. 
The factorization scale is associated with the typical momentum transfer of the relevant physical process. For the partonic picture to be valid, the factorization scale must be significantly larger than $\Lambda_{\rm QCD}$. In our case, the $Z'$ is produced on-shell, so we thus take $Q^2= m_{Z'}^2>(2\ {\rm GeV})^2$. As presented 
in Ref.~\cite{Burk:2026mvk}, the cross section for $NN$ production via $Z'$ production and decay in $hp$ collisions is given by 
\begin{eqnarray}
\label{eq:NNxs}
\sigma(h p \to Z' \to NN) &=& \frac{g_{Z'}^4}{1944\, \Gamma_{Z'}}\frac{\beta^3_N m_{Z'}}{s}  \nonumber \\
&&\times \frac{1}{2}\sum_{q,\bar{q}}\int_{m_{Z'}^2/s}^1\frac{dx_2}{x_2}\left[f_{q/h}\left(\frac{m_{Z'}^2}{x_2s}, m_{Z'}^2\right)f_{\bar{q}/p}(x_2, m_{Z'}^2) + (q\leftrightarrow \bar{q})\right] ,
\end{eqnarray}
where $\beta_N=(1-4m_N^2/\hat{s})^{1/2}$ and $\Gamma_{Z'}$ is the $Z'$ total width. To simulate our Drell-Yan production, we modified the \texttt{B-L-4UFO} \cite{Basso:2008iv, Deppisch:2018eth,Amrith:2018yfb} model file in \texttt{FeynRules}~\cite{Alloul:2013bka} to allow arbitrary couplings to an HNL via $U_e, U_\mu, U_\tau$, and performed our MC event generation using \texttt{WHIZARD}~\cite{Kilian:2007gr}.

\subsection{HNLs From ATLAS Interaction Point and Detection Criteria}
\label{subsec:ATLASIP}

\begin{figure}[tb]
    \centering 
\includegraphics[width=0.465\linewidth]{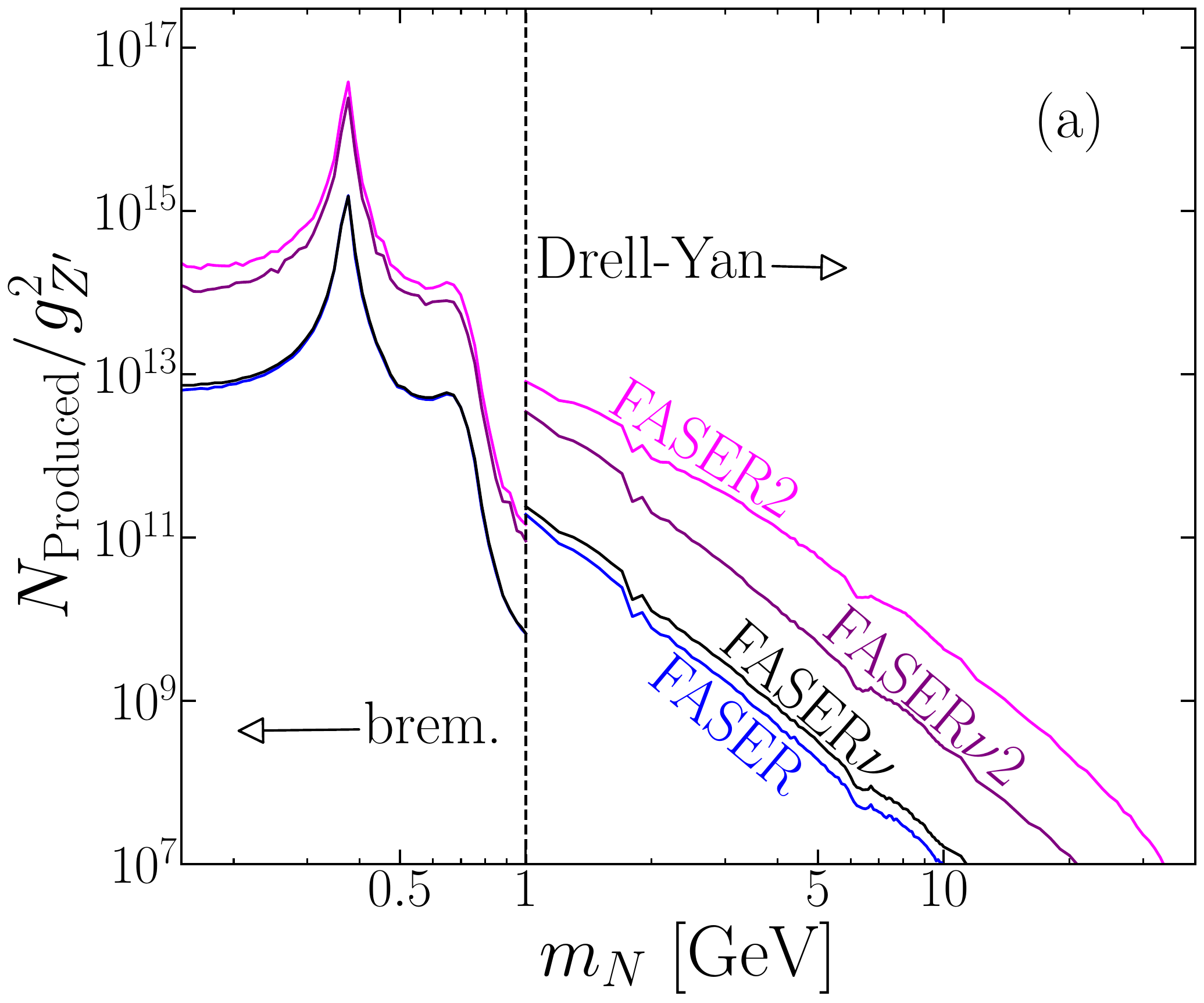}
 \\
 \includegraphics[width=0.475\linewidth]{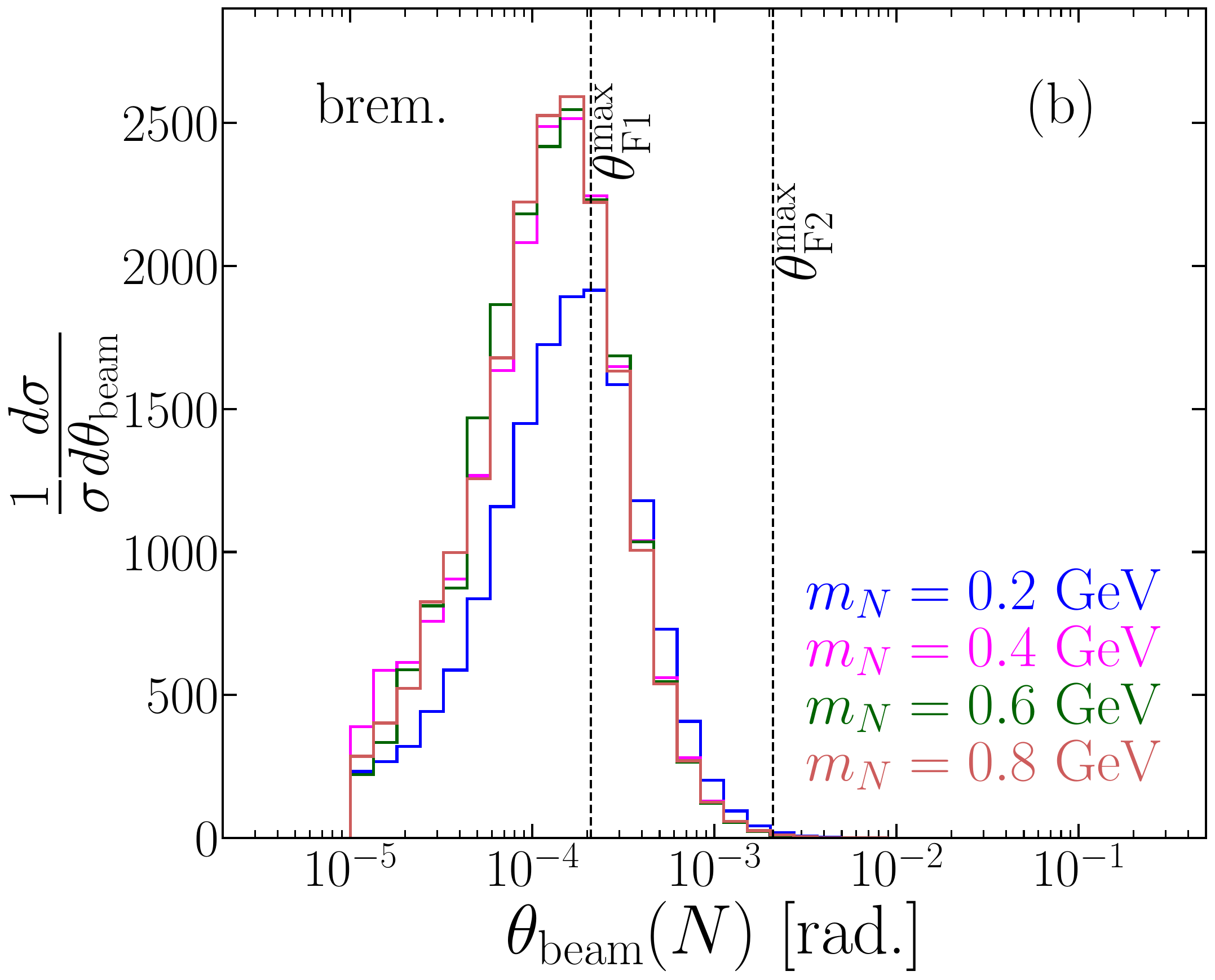}\includegraphics[width=0.465\linewidth]{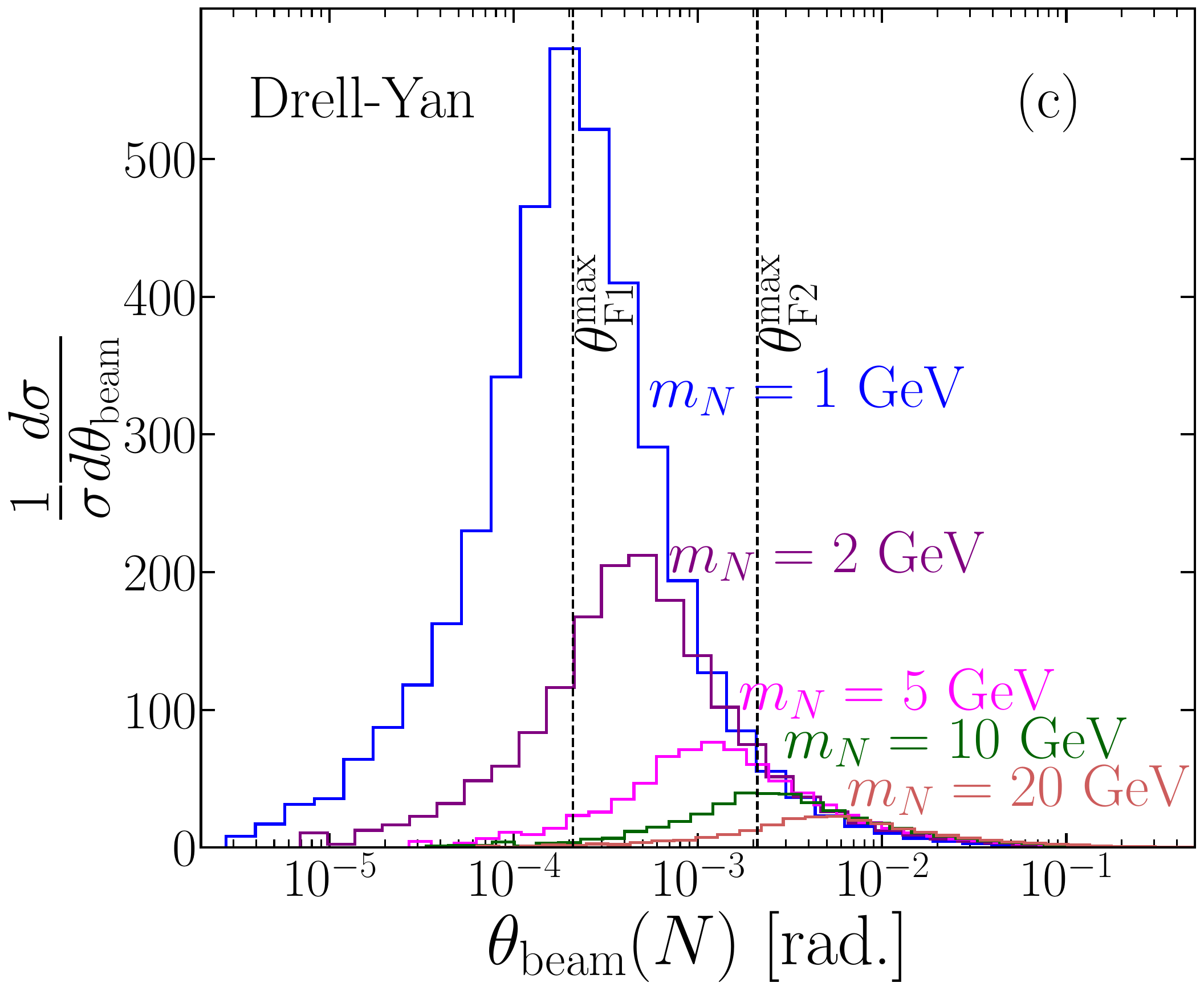}
\centering
   \caption{ (a) HNL fluxes normalized to the squared gauge coupling within the geometric acceptance of each benchmark experiment, including proton bremsstrahlung and Drell-Yan production modes. The vertical dashed line shows the transition from bremsstrahlung to Drell-Yan domination. (b,c) Beam angle distributions of HNLs produced from proton bremsstrahlung and Drell-Yan for various masses $m_N$. The maximum geometrical acceptances from the ATLAS IP to FASER and FASER2 are displayed by the vertical dashed lines. Here we have fixed $m_{Z'}/m_N=2.1$.}
\label{Fig.HNL_eff_and_Flux}
\end{figure} 
During Run 3, the LHC operated at COM energy $13.6$ TeV. For the HL-LHC, the COM energy is expected to increase to $14$ TeV. We found that the difference in sensitivities between these energies is negligible. For simplicity, we take $\sqrt{s} = 14$ TeV for both and luminosities as shown in Table~\ref{tab:expsummary}. 
It is important to note that FASER and FASER2 have rather different geometric angular acceptances in the forward region~\cite{FASER:2018bac,FPF:2025bor}. In our analysis, we adopt 
\begin{equation}
    \theta^{\text{max}}_{\text{F1}} \sim 2\times 10^{-4}\ {\rm rad\ for \ FASER},\qquad  
     \theta_{\text{F2}}^{\text{max}} \sim 2\times 10^{-3}\ {\rm rad\ for \ FASER2}, 
     \label{eq:angle}
\end{equation}
based on the detector geometry and the baseline discussed in Section~\ref{sec:Exp_Settings}.

As described in the previous subsections, HNLs may be produced via the production and decay of $Z'$ at the ATLAS IP. For the sake of illustration, we take $m_{Z'} = 2.1 m_N$, to ensure a prompt decay $Z' \to NN$. 
We include contributions that arise both from proton bremsstrahlung and also from perturbative QCD production.

Taking the bremsstrahlung production to be for $m_{Z'} \lesssim 2$ GeV and the Drell-Yan production mode to be for $m_{Z'} \gtrsim 2$ GeV, 
in Fig.~\ref{Fig.HNL_eff_and_Flux}(a), we present the HNL flux for each experiment within the geometric acceptance of Eq.~(\ref{eq:angle}). We first note that the contribution from proton bremsstrahlung dies out for $m_{Z'}\ge 1$ GeV due to the nucleon form factor suppression, and those from Drell-Yan take over at higher momentum transfer.

We observe 
that FASER2 and FASER$\nu$2 may collect substantially more HNLs, which is primarily due to the larger integrated luminosity of the HL-LHC and the larger angular acceptance. This is evidenced by Figs.~\ref{Fig.HNL_eff_and_Flux}(b) and \ref{Fig.HNL_eff_and_Flux}(c), where we show the HNL's angular distribution in the laboratory frame with respect to the proton beam direction for proton bremsstrahlung and Drell-Yan production modes, respectively, at various masses. In particular, we note that the HNLs produced from bremsstrahlung are very forward, with the majority of events landing within FASER's acceptance. On the other hand, the HNLs produced from Drell-Yan have a larger angular spread, which increases with the HNL mass thus resulting in fewer signal events.

Once the produced HNLs reach a detector of interest as specified in Eq.~(\ref{eq:eff}), they decay into some visible final states within the detector geometry. Various HNL decay channels are available. Among them, we consider the following signal channels for our final state HNLs:
\begin{itemize}
    \item  $N \to \nu\ell^+  \ell^-$, \quad $\ell\in \{ e, \mu\}$
    \item $N \to \nu\pi^0$, \quad $\pi^0 \to \gamma\gamma$,
    \item $N\to \ell^{\pm} + \text{ hadrons}$, \quad $\ell\in \{e, \mu, \tau\}$
\end{itemize}

With suitable cuts on the final-state particle energies, the backgrounds in all four experiments can be reduced to negligible levels. At FASER, various backgrounds arise from high-energy muons or neutrinos coming from the ATLAS IP and from instrumental effects. Recently, the FASER Collaboration has presented an analysis on dark photons providing a two-track signal, in which they accounted for potential backgrounds associated with veto inefficiencies, neutral hadrons, muons missing the veto, neutrinos, and non-collision background~\cite{FASER:2023tle}.  The study suggests that the backgrounds for the charged multi-track signals should be negligible, provided that we take the energy cuts~\cite{Feng:2024zfe}:  
\begin{equation}
E_{e^+e^-} > 100 \text{ GeV},  \qquad E_{\mu^+\mu^-} > 100 \text{ GeV},
\end{equation} 
where $E_{\ell^+\ell^-}=E_{\ell^+}+E_{\ell^-}$.
For the diphoton final states, there is a potentially large background from neutrinos interacting with the end of the FASER detector, especially for energies below 1 TeV. The FASER Collaboration has planned a high-precision preshower upgrade to address the sizable diphoton background from neutrinos, which may resolve pairs of photons that pass the cuts~\cite{Boyd:2803084}:
\begin{equation}
    E_{\gamma} > 100 \text{ GeV}, \qquad \Delta_{\gamma\gamma} \gtrsim 200 \text{ $\mu$m},
\end{equation}
where $\Delta_{\gamma\gamma}$ denotes the transverse separation between the two photons at the preshower plane and we compute it by taking $\Delta_{\gamma\gamma} \sim L\times(2m_{\pi^0}/E_{\pi^0})$, where $L$ is the length from the $\pi^0$ decay to the back end of the FASER spectrometer. Regarding this selection, we assume an event identification efficiency of $80\%$.
For the remaining channel for FASER, i.e., the $\ell^\pm +{\rm hadrons}$ final states, no dedicated background studies are available. We thus simply adopt~\cite{FASER:2018eoc}:
\begin{equation}
    E_{\text{vis.}} > 100 \text{ GeV} \, .
\end{equation}

When it comes to FASER2, although its design has not yet been finalized, we expect that it will have similar or better background rejection capabilities as the currently operating FASER. Therefore, we assume that the same search strategies will enable us to perform background-free analyses as for FASER. 

For FASER$\nu$, the final-state particles are determined by looking for charged tracks using an emulsion detector~\cite{FASER:2020gpr}. As such, we do not consider the diphoton final states for FASER$\nu$ and its proposed upgrade. Electron events are detected by reconstructing the electromagnetic showers through the emulsion layers. Muon events are detected by measuring the muon's momentum deposit at the FASER spectrometer. The excellent spatial resolution of the emulsion films allows for studying the tau lepton tracks to record tau events. Using the latest available results from the FASER Collaboration we take \cite{FASER:2026jhh, FASER:2019dxq}
\begin{equation}
E_{e^{\pm}} > 200 \text{ GeV},\qquad E_{\mu^{\pm}} > 100\text{ GeV}, \qquad E_{\tau^{\pm}} > 100 \text{ GeV}
\end{equation}
for the final sensitivity estimates at FASER$\nu$. We apply lepton identification efficiencies of $70\%$ for the electron flavor~\cite{FASER:2026jhh}, $86\%$ for the muon flavor, and $75\%$ for the tau flavor~\cite{Kling:2023tgr}. Although FASER$\nu$2 will have considerably upgraded energy resolution and likely better detection efficiencies than FASER$\nu$, we conservatively take the same efficiency estimates and cuts as above for our analysis. 

\subsection{Sensitivity Reach}
\label{sec:Sensitivity}

For the sake of illustration, we fix the following benchmark gauge couplings for the rest of the presentation for each production mode to be consistent with model-dependent bounds~\cite{Ilten:2018crw, Bauer:2018onh}:  
\begin{equation}
    g^{\text{brem.}}_{B-L} = 10^{-4},
    \qquad g^{\text{brem.}}_{B-3L_{\tau}} = 10^{-4},
    \qquad 
    g_{B}^{\text{brem.}} = 10^{-3}. 
    \label{eq:coup_brem}
\end{equation}
\begin{equation}
    g^{\text{DY}}_{B-L} = 10^{-4},\qquad 
    g^{\text{DY}}_{B-3L_\tau} = 10^{-3},\qquad 
    g_{B}^{\text{DY}} = 10^{-2}. 
    \label{eq:coup_DY}
\end{equation} 
Furthermore, we take a fixed mass ratio  $m_{Z'}/m_N=2.1$.  

\begin{figure} [tb]
    \centering
    \includegraphics[width=1.\linewidth]{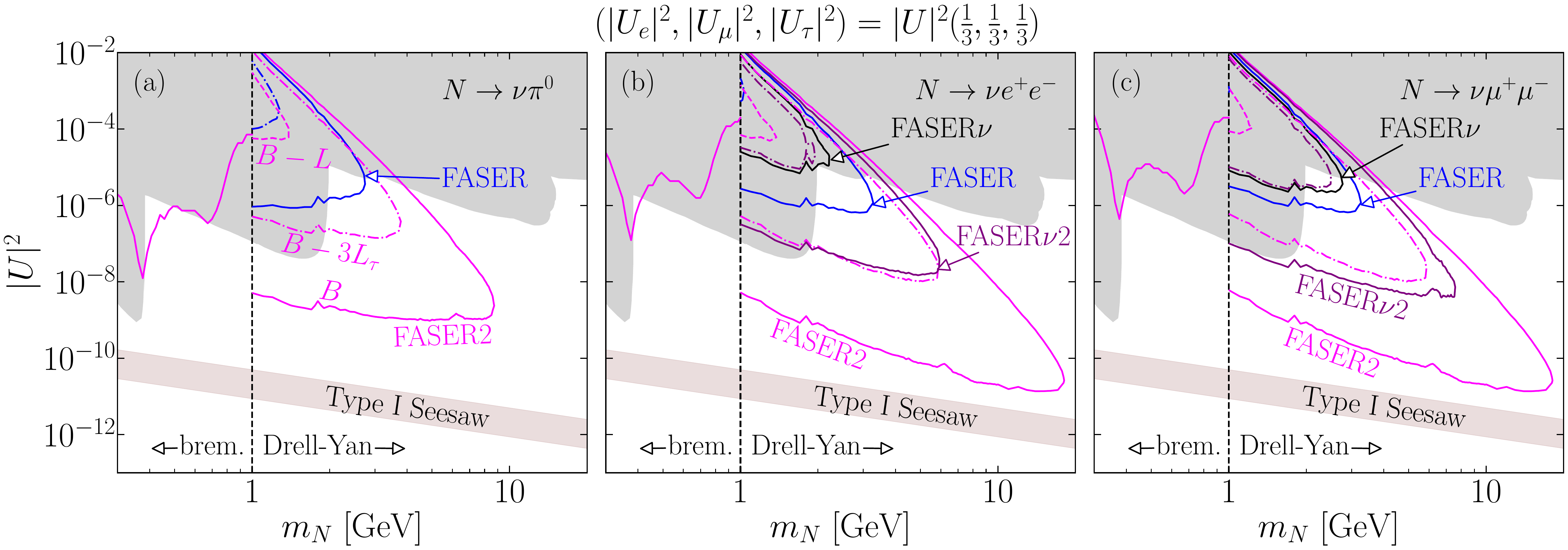}

    \includegraphics[width=1.\linewidth]{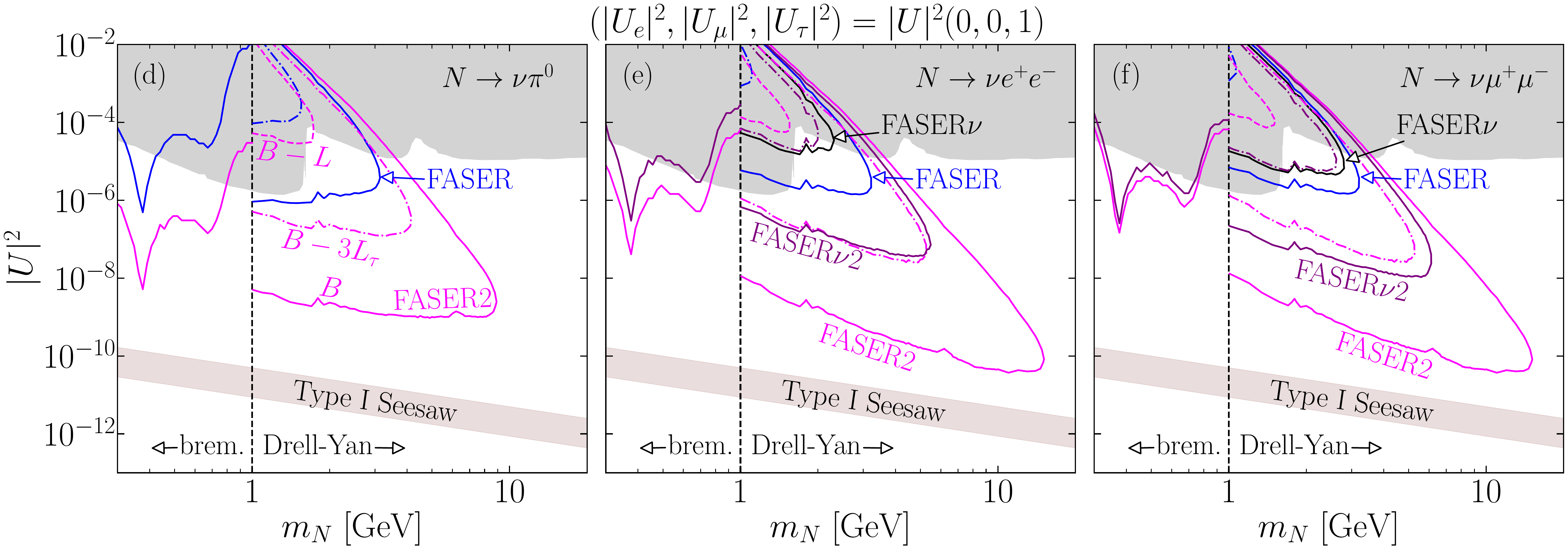}
    \caption{$90\%$ C.L. HNL sensitivity contours in the mass-mixing plane for the $N\to \nu \pi^0$ (left), $N\to \nu e^+e^-$ (middle), and $N\to \nu\mu^{+}\mu^-$ (right) final states taking neutrino flavor mixing benchmark $(U_e^2, U_{\mu}^2, U^2_{\tau}) = U^2(\frac{1}{3}, \frac{1}{3}, \frac{1}{3})$ (upper) and $U^2(0, 0, 1)$ (lower) for the FASER (blue), FASER$\nu$ (black), FASER2 (magenta), and FASER$\nu$2 (purple) experiments. The dashed, dot-dashed, and solid curves are for the $U(1)_{B-L}$, $U(1)_{B-3L_{\tau}}$, and $U(1)_B$ models, respectively, for the appropriate benchmark couplings in Eqs.~(\ref{eq:coup_brem}) and ~(\ref{eq:coup_DY}). The upper gray-shaded region represents the model-independent HNL mass-mixing limits. The lightly shaded band in the bottom half plane shows the canonical Type-I seesaw prediction to get the correct atmospheric (upper line) and solar (lower line) mass-squared splittings. The vertical line shows the transition from the bremsstrahlung to the Drell-Yan dominated production regime. Here we have fixed the mass ratio $m_{Z'}/m_N = 2.1$.}
    \label{Fig.HNL_fluxnuXevents}
\end{figure}

\begin{figure} [tb]
    \centering
    \includegraphics[width=1\linewidth]{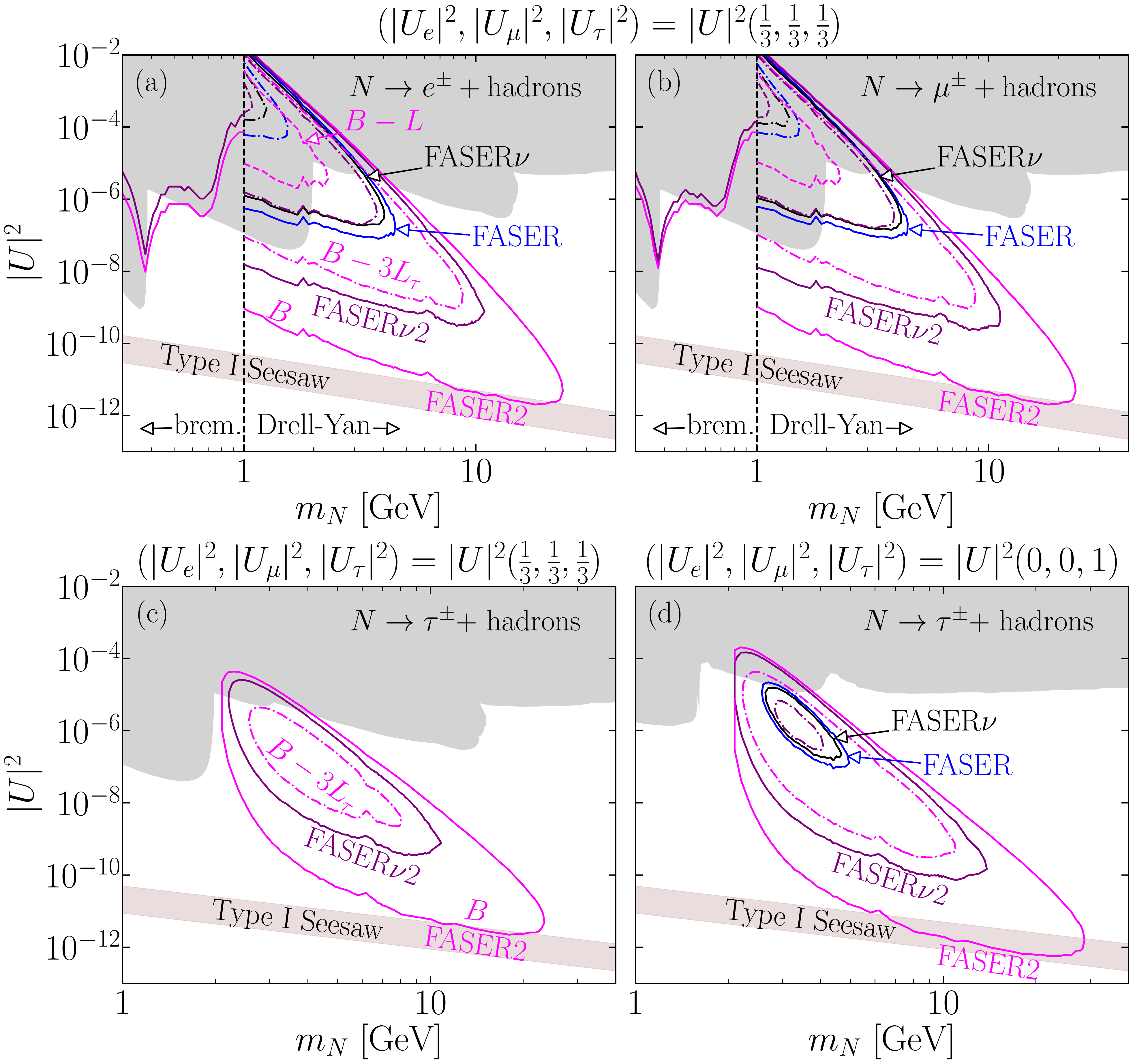}
    \caption{
    Same as in Fig.~\ref{Fig.HNL_fluxnuXevents} but for the $\ell^\pm$+hadrons final state.
    }\label{Fig.HNL_fluxlhadronsevents}
\end{figure}

After applying the optimal cuts, we present in Figs.~\ref{Fig.HNL_fluxnuXevents} and \ref{Fig.HNL_fluxlhadronsevents} the resulting HNL signal sensitivities at the 90\% C.L. in the $|U|^2\text{-}m_N$ plane for the four experiments and the three benchmark models discussed above. The magnitude difference among the models is primarily due to the benchmark choice of the couplings as in Eqs.~(\ref{eq:coup_brem}) and (\ref{eq:coup_DY}), although there are also minor model-dependent factors. For both figures, we present sensitivities for neutrino flavor mixing benchmarks given by Eqs.~\eqref{eq:111} and \eqref{eq:001}. For the flavor-democratic benchmark, we recast the current bounds as described in Section~\ref{sec:model}. For the $\tau$-specific benchmark, the upper gray-shaded region represents the model-independent neutrino mass-mixing limits coming from DELPHI~\cite{DELPHI:1996qcc}, CHARM~\cite{Boiarska:2021yho}, BEBC~\cite{Barouki:2022bkt} and ArgoNeuT~\cite{ArgoNeuT:2021clc}. The lower light-brown band is the target region predicted by the canonical Type-I Seesaw mass mixing relation $|U|^2\sim m_{\nu}/m_N$~\cite{Minkowski:1977sc, Mohapatra:1979ia, Yanagida:1979as, Gell-Mann:1979vob}, where we have taken $\sqrt{\Delta m^2_{\rm atm}}\simeq 50$ meV (atmospheric) and $\sqrt{\Delta m^2_{\rm sol}}\simeq 8.7$ meV (solar)~\cite{Esteban:2024eli} 
as the representative values for $m_\nu$ to get the upper and lower edges of the band, respectively.

Starting with Fig.~\ref{Fig.HNL_fluxnuXevents}, we present the sensitivity of each experiment to HNLs produced via bremsstrahlung and Drell-Yan processes decaying via $N\to \nu\pi^0$ (left), $N\to \nu e^+e^-$ (middle), and $N\to \nu \mu^+\mu^-$ (right) final states at the $90\%$ C.L. in the $|U|^2 -m_N$ plane for $111$ (top) and $001$ (bottom) neutrino flavor mixing benchmarks. For the HNLs produced via bremsstrahlung, we only present the sensitivities taking the $U(1)_B$ model benchmark, as the sensitivities for other model benchmarks remain completely within the already excluded regions of parameter space. For the bremsstrahlung production, we find that the $\nu \pi^0$ channel presented in Figs.~\ref{Fig.HNL_fluxnuXevents}(a) and \ref{Fig.HNL_fluxnuXevents}(d) is the most sensitive channel, reaching peak sensitivity for HNL mass $m_N\approx 0.4$ GeV and $|U|^2\approx 10^{-6}$ ($10^{-8}$) for FASER (FASER2) in the 001 (111 and 001) flavor benchmark. As for the Drell-Yan production, the $\nu\pi^0$ channel is also sensitive to new regions of parameter space, with FASER's $U(1)_B$ benchmark reaching $|U|^2\approx 10^{-6}$ up to $m_N\approx 4$ GeV for both neutrino mixing patterns. FASER2 achieves coverage across all three model benchmarks, reaching $|U|^2\approx 10^{-4}$ for $m_N\approx 2$ GeV in $U(1)_{B-L}$, $|U|^2\approx 10^{-7}$ for $1 \text{ GeV} \lesssim m_N\lesssim 4 \text{ GeV}$ for $U(1)_{B-3L_{\tau}}$, and $|U|^2\approx 10^{-8}$ for  $1 \text{ GeV} \lesssim m_N\lesssim 10 \text{ GeV}$ for $U(1)_B$ across both neutrino mixings. 

Turning now to Figs.~\ref{Fig.HNL_fluxnuXevents}(b) and \ref{Fig.HNL_fluxnuXevents}(c), we observe that FASER$\nu$2 may probe coupling $|U|^2\approx 10^{-6}$ for mass $m_N\approx 0.4$ GeV, and FASER2 may reach as low $|U|^2\approx 3\times 10^{-7}$ for $m_N\approx 0.4$ in both neutrino mixing scenarios. All four experiments may reach unexcluded phase space in this channel too, with FASER reaching $|U|^2\approx 10^{-6}$ for $1 \text{ GeV} \lesssim m_N \lesssim 4 \text{ GeV}$ in the $U(1)_B$ model with both neutrino mixings. FASER$\nu$ is also sensitive to unexcluded parameter space in this channel, reaching $|U|^2\approx 10^{-5}$ for $m_N\approx 2$ GeV for the $U(1)_B$ model benchmark with 001 neutrino mixing. Next, FASER$\nu$2 reaches sensitivities $|U|^2\approx 10^{-5}$ for masses $m_N\approx 2$ GeV in the $U(1)_{B-3L_{\tau}}$ model benchmark, as well as $|U|^2\approx 10^{-7}$ for masses $1 \text{ GeV} \lesssim m_N\lesssim 6 \text{ GeV}$ in the $U(1)_B$ model benchmark for both masses and neutrino mixings. FASER2 may achieve an even better reach, extending to $|U|^2\approx 10^{-8}$ near $m_N\approx 6$ GeV for the $U(1)_{B-3L_{\tau}}$ model, and probing as low as $|U|^2\approx 10^{-10}$ for masses $m_N\approx 20$ GeV for both neutrino mixings. Finally, the $\nu\mu^+\mu^-$ channel in Figs.~\ref{Fig.HNL_fluxnuXevents}(c) and \ref{Fig.HNL_fluxnuXevents}(f) may reach slightly stronger sensitivities that $\nu e^+ e^-$ due to the less stringent energy cuts for FASER and FASER2, and the higher identification efficiencies for FASER$\nu$ and FASER$\nu$2. Overall, we observe that FASER, FASER$\nu$2, and FASER2 are sensitive to HNLs produced via proton bremsstrahlung, and all four experiments may be sensitive to HNLs produced from Drell-Yan for the appropriate model benchmarks.

Moving now to Fig.~\ref{Fig.HNL_fluxlhadronsevents}, we present the sensitivity of each experiment at the $90\%$ C.L. in the $|U|^2 -m_N$ plane to HNLs produced via proton bremsstrahlung and Drell-Yan decaying via (a) $N\to e^{\pm} + \text{hadrons}$, (b) $N\to \mu^{\pm} + \text{hadrons}$, and (c) $N\to \tau^{\pm} + \text{hadrons}$ for the $111$ neutrino flavor mixing benchmark, and also (d) for $N\to \tau^{\pm} + \text{hadrons}$ for the $001$ neutrino flavor mixing benchmark. For Figs.~\ref{Fig.HNL_fluxlhadronsevents}(a) and ~\ref{Fig.HNL_fluxlhadronsevents}(b), we model the bremsstrahlung production final states as $\ell^{\pm} + \pi^{\mp}$. In both cases, these channels are sensitive to couplings $|U|^2\approx 10^{-8}$ for $m_N\approx 0.4$ GeV for both FASER$\nu$2 and FASER2 in the $U(1)_B$ benchmark. As for the Drell-Yan production for both Figs.~\ref{Fig.HNL_fluxlhadronsevents}(a) and ~\ref{Fig.HNL_fluxlhadronsevents}(b), FASER and FASER$\nu$ are both sensitive to $|U|^2\approx 5\times 10^{-7}$ for $m_N\approx 5$ GeV in the $U(1)_B$ model, FASER$\nu$2 is sensitive to $|U|^2\approx 10^{-9}$ for $m_N\approx 10$ GeV for the $U(1)_B$ model, and FASER2 may achieve sensitivities of $|U|^2\approx 10^{-11}$ for $m_N\approx 20$ GeV for the $U(1)_B$ model. From Fig.~\ref{Fig.HNL_fluxlhadronsevents}(c), we observe that FASER$\nu$2 is sensitive to $|U|^2\approx 10^{-9}$ for $m_N\approx 10$ GeV in the $U(1)_B$ model benchmark and FASER2 achieves sensitivities down to $|U|^2\approx 10^{-11}$ for masses $m_N\approx 20$ GeV in the $U(1)_B$ model benchmark with 111 neutrino flavor mixing. For 001 neutrino flavor mixing as shown in Fig.~\ref{Fig.HNL_fluxlhadronsevents}(d), the sensitivities are substantially better, with FASER and FASER$\nu$ both being sensitive to couplings $|U|^2\approx 10^{-7}$ for masses $m_N\approx 5$ GeV in the $U(1)_B$ benchmark, FASER$\nu$2 being sensitive to $|U|^2\approx 10^{-10}$ for masses $m_N\approx 15$ GeV in the $U(1)_B$ benchmark, and finally FASER2 reaching couplings as low as $|U|^2\approx 10^{-12}$ for $10 \text{ GeV} \lesssim m_N\lesssim 30\text{ GeV}$. Overall, we find that FASER$\nu$2 and FASER2 are sensitive to HNLs from proton bremsstrahlung, and all four experiments are sensitive to lepton and hadrons final states for both neutrino flavor mixing benchmarks. In particular, in all three scenarios, the substantial branching ratio for $N\to \ell^{\pm} + \text{hadrons}$ allows the FASER2 to reach the Type-I Seesaw band for all lepton flavors for HNL masses $10 \text{ GeV} \lesssim m_N\lesssim 20$ GeV for the 111 neutrino mixing scenario, and to reach the Type-I Seesaw band for HNL masses $6\text{ GeV}\lesssim m_N \lesssim 30\text{ GeV}$ with the 001 neutrino mixing.

\begin{figure} [tb]
    \centering
    \includegraphics[width=0.9\linewidth]{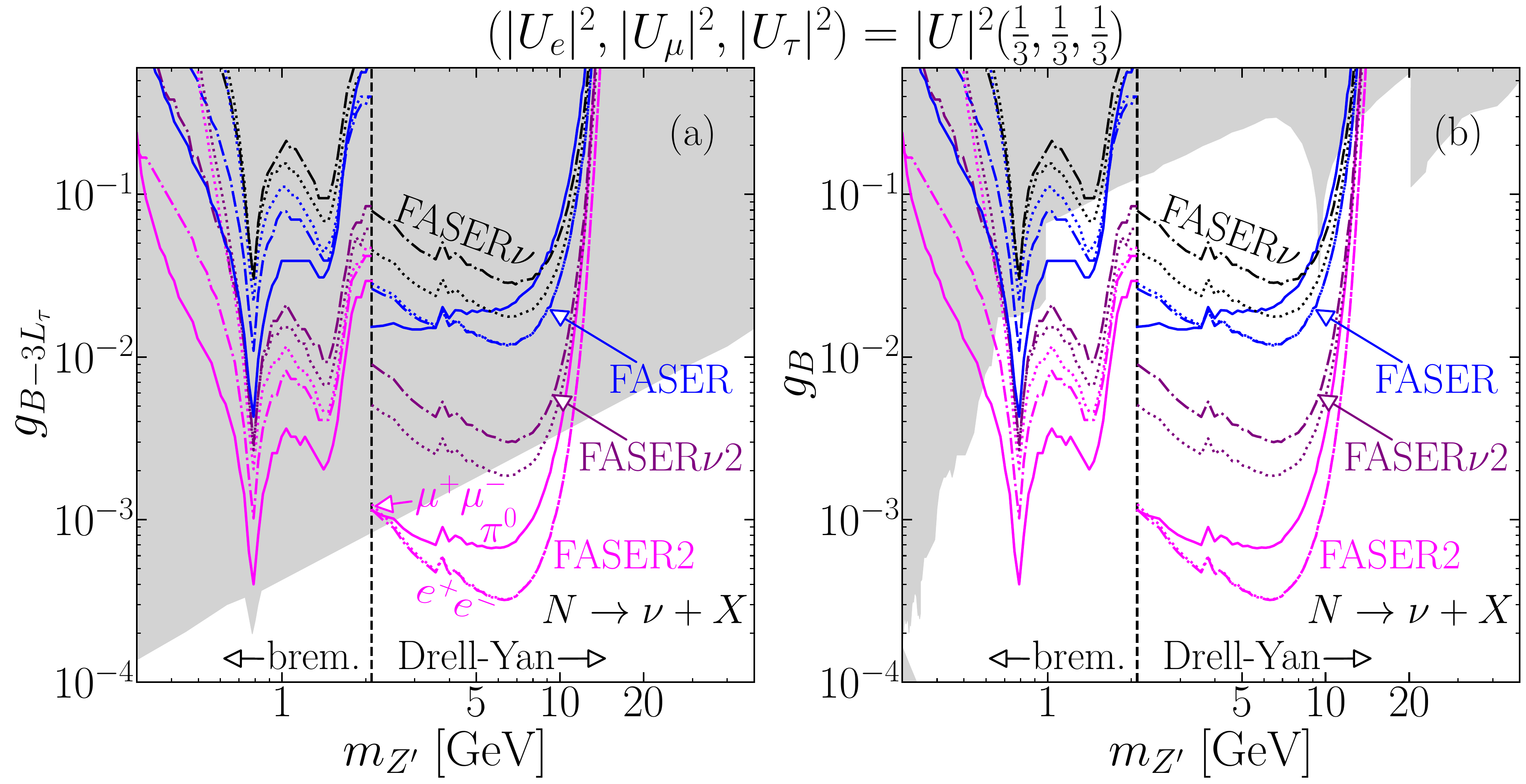}
    \includegraphics[width=0.9\linewidth]{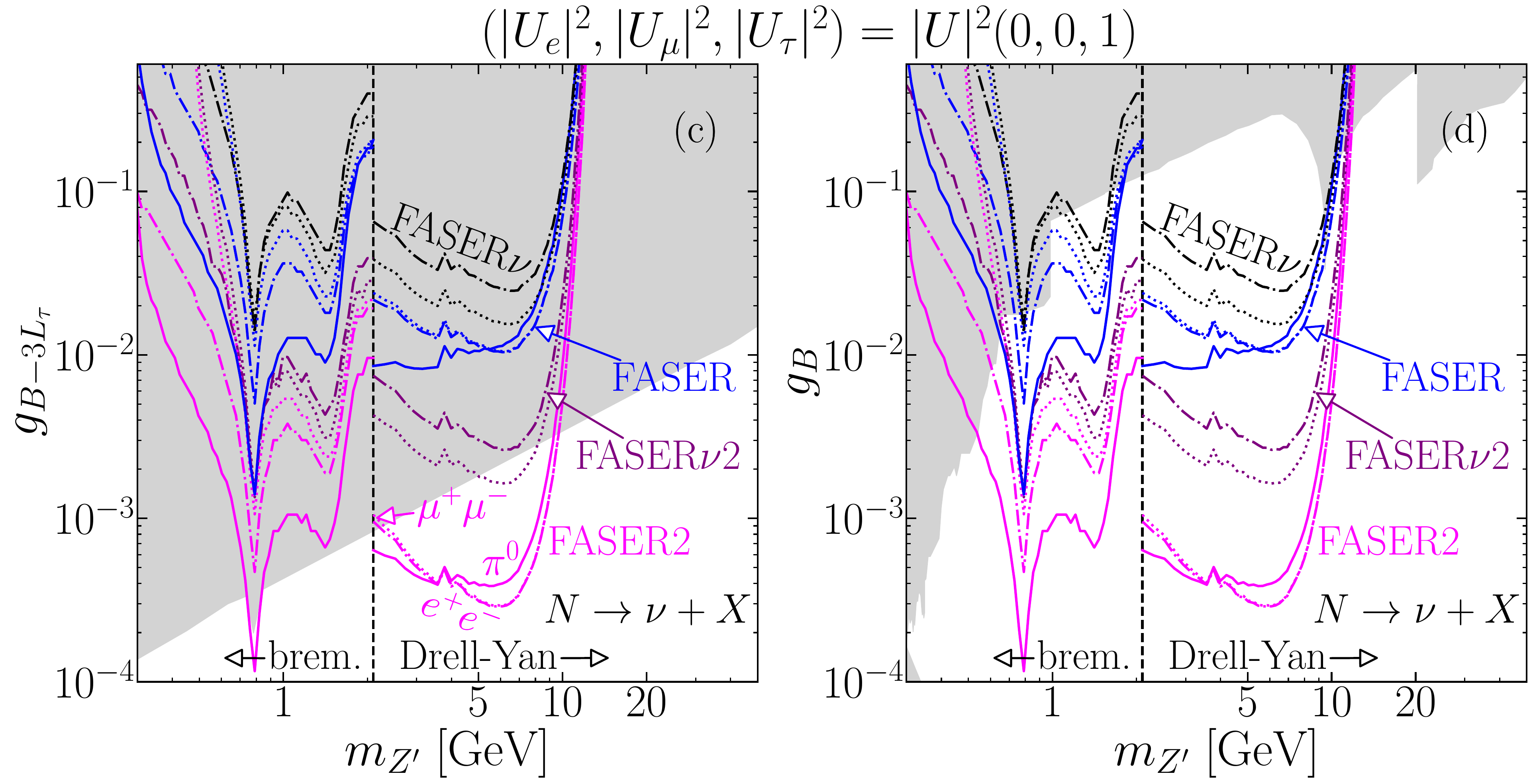}
    
    \caption{90\% C.L. sensitivities in the gauge boson mass-coupling plane at FPF facilities considering the HNL production via bremsstrahlung and Drell-Yan and decay into $\nu+ X$ final states for the benchmark models $U(1)_{B}$ and $U(1)_{B - 3L_{\tau}}$ for the 111 (top) and 001 (bottom) neutrino flavor mixing benchmarks. We assume a mass ratio $m_{Z'}/m_N = 2.1$. The gray-shaded region shows the existing exclusion limits. }
    \label{fig:gauge_nuX}
\end{figure}

\begin{figure} [tb]
    \centering
    \includegraphics[width=0.9\linewidth]{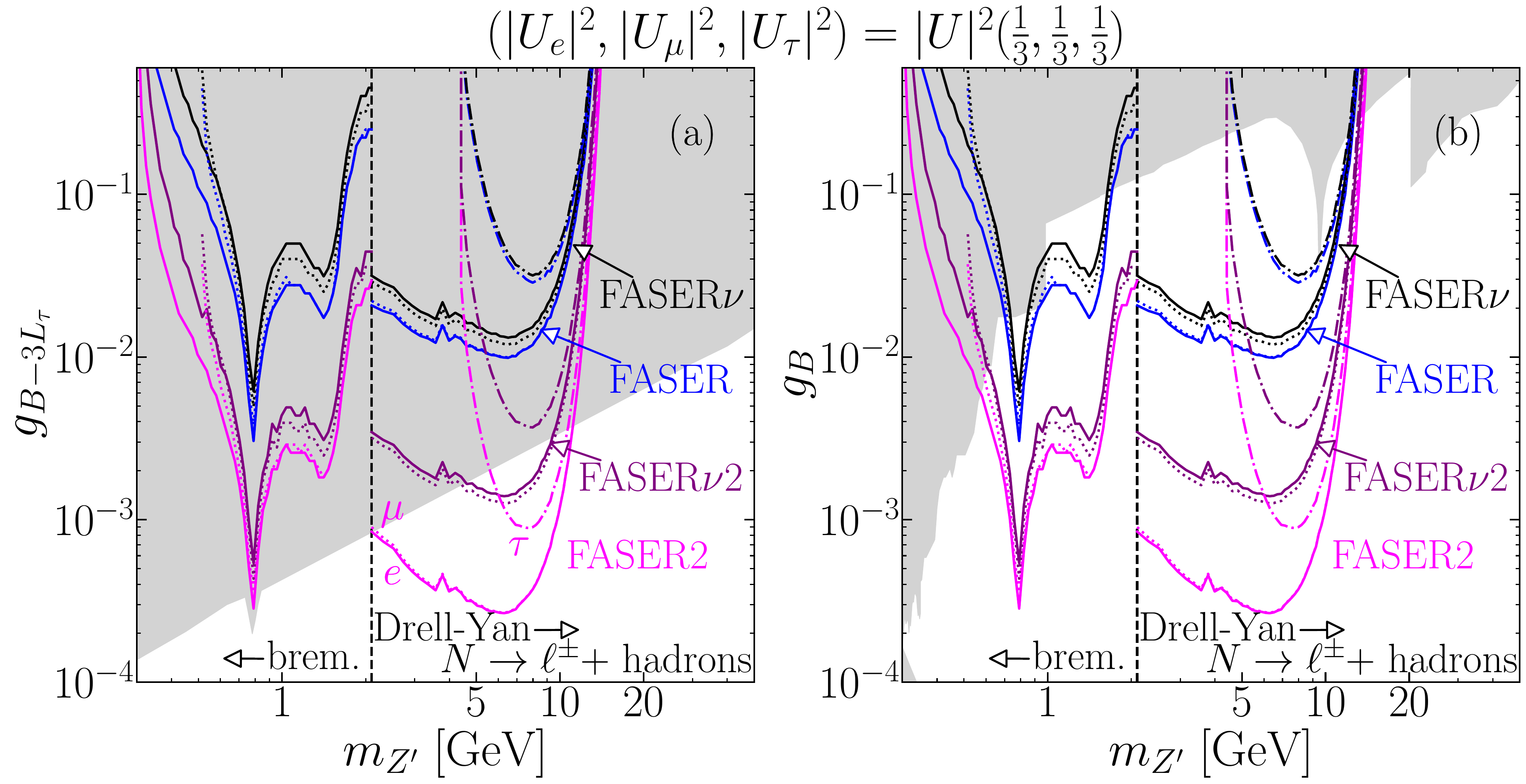}
    \includegraphics[width=0.9\linewidth]{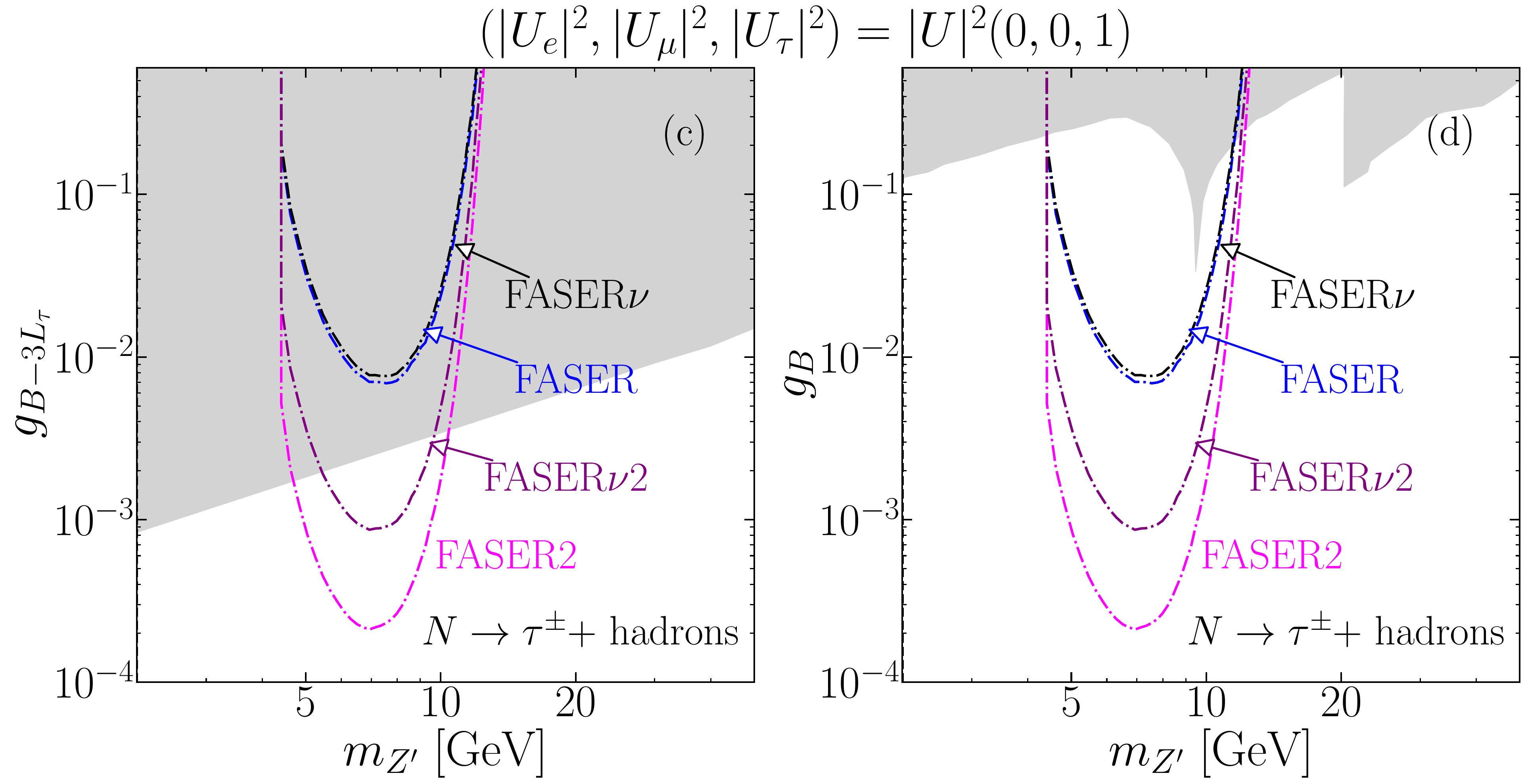}
    \caption{Same as in Fig.~\ref{fig:gauge_nuX} but for the $\ell^\pm$+hadrons final state.}
    \label{fig:lhadrons_gaugeSensitivities}
\end{figure}

In Fig.~\ref{fig:gauge_nuX}, we present the projected $Z'$ sensitivity reach at $90\%$ C.L. in the $g_{B-3L_{\tau}}$~–~$m_{Z'}$ (left) and $g_{B}$~–~$m_{Z'}$ (right) planes for the  $N\to \nu\pi^0$ (solid), $N\to \nu e^+e^-$ (dashed), and $N\to \nu e^+e^-$ (dotted) channels at FASER, FASER$\nu$, FASER2, and FASER$\nu$2 taking benchmark coupling $|U|^2 = 10^{-7}$ for the 111 flavor benchmark (top), and $|U|^2 = 10^{-6}$ for the 001 flavor benchmark (bottom). Overall, we do not observe a significant difference in sensitivities between models, because the models only differ by minor model-dependent factors. For Figs.~\ref{fig:gauge_nuX}(a) and \ref{fig:gauge_nuX}(c), we observe that FASER$\nu$2 may probe the parameter space using the $N\to \nu\mu^+\mu^-$ channel down to $g_{B-3L_{\tau}}\approx 2\times10^{-3}$ for $m_{Z'} \approx 7$ GeV. Additionally, FASER2 may probe the $N\to \nu \pi^-$ channel down to $g_{B-3L_{\tau}}\approx 8\times 10^{-4}$ for $m_{Z'}\approx 6$ GeV in the 111 flavor benchmark, and lower to $g_{B-3L_{\tau}}\approx 10^{-4}$ for $m_{Z'}\approx 0.8$ GeV and $g_{B-3L_{\tau}}\approx 3\times 10^{-4}$ for $m_{Z'}\approx 6$ GeV in the 001 flavor benchmark. In both flavor benchmarks, FASER2 may probe $N\to \nu e^+e^-$ and $N\to \nu\mu^+\mu^-$ down to $g_{B-3L_{\tau}}\approx 3\times 10^{-4}$ near $m_{Z'}\approx 6$ GeV. For the $U(1)_B$ model benchmarks, FASER, FASER$\nu$, and FASER$\nu$2 may also probe outside of the excluded parameter space. In particular, because of the higher muon identification efficiency at FASER$\nu$ and FASER$\nu$2, the channel $N\to \nu\mu^+\mu^-$ is more sensitive than $N\to \nu e^+ e^-$, with FASER$\nu$ (FASER$\nu$2) reaching couplings as low as $g_B\approx 2\times 10^{-2}$ ($2\times 10^{-3}$) near both $m_{Z'}\approx 0.8$ GeV and $m_{Z'}\approx 6$ GeV for both flavor benchmarks. 

In Fig.~\ref{fig:lhadrons_gaugeSensitivities}, we present the projected $Z'$ sensitivity reach at the $90\%$ C.L. in the $g_{B-3L_{\tau}}$~–~$m_{Z'}$ (left) and $g_{B}$~–~$m_{Z'}$ (right) planes for the  $N\to e^{\pm} + \text{hadrons}$ (solid), $N\to \mu^{\pm} + \text{hadrons}$ (dotted), and $N\to \tau^{\pm} + \text{hadrons}$ (dashed) channels at FASER, FASER$\nu$, FASER2, and FASER$\nu$2 taking benchmark coupling $|U|^2 = 10^{-7}$ for the 111 flavor benchmark (top), and $|U|^2 = 10^{-6}$ for the 001 flavor benchmark (bottom). Again, we do not observe a significant difference in sensitivities between models, because the models only differ by minor model-dependent factors. From Figs.~\ref{fig:lhadrons_gaugeSensitivities}(a) and \ref{fig:lhadrons_gaugeSensitivities}(c), we observe that FASER$\nu$2 may probe the mass-coupling parameter space using the $N\to e^{\pm} + \text{hadrons}$ and $ N\to \mu^{\pm} + \text{hadrons}$ channels down to $g_{B-3L_{\tau}}\approx 10^{-3}$ for $5 \text{ GeV}\lesssim m_{Z'}\lesssim 10 \text{ GeV}$ for the 111 benchmark, as well as the same mass-coupling for the $N\to \tau^{\pm} + \text{hadrons}$ channel in the 001 benchmark. For the $U(1)_B$ model in Figs.~\ref{fig:lhadrons_gaugeSensitivities}(b) and \ref{fig:lhadrons_gaugeSensitivities}(d), all four experiments may probe a large amount of the parameter space through all channels, with FASER and FASER$\nu$ both being sensitive to as low as $g_{B}\approx 5\times 10^{-3}$ for $m_{Z'}\approx 0.8$ GeV in the $N\to e^{\pm} + \text{hadrons}$ and $N\to \mu^{\pm} + \text{hadrons}$ channels in the 111 benchmark, and sensitive to $g_{B}\approx 10^{-2}$ for $m_{Z'} \approx 7$ GeV for the $N\to \tau^{\pm} + \text{hadrons}$ channel in the 001 benchmark. FASER2 (FASER$\nu$2) fare even better, reaching $g_{B}\approx 5\times 10^{-4} ~(10^{-4})$ at $m_{Z'}\approx 0.8$ GeV (0.8 GeV) for the $N\to e^{\pm} + \text{hadrons}$ and $N\to \mu^{\pm} + \text{hadrons}$ channels in the 111 benchmark, and $g_{B}\approx 2\times 10^{-4}~(10^{-3})$ for the $N\to \tau^{\pm} + \text{hadrons}$ channel in the 001 benchmark. 

\section{Summary and Conclusion}
\label{sec:conclusion}

In this work, we have investigated the prospects for HNL searches in the far-forward LHC experiments FASER and FASER$\nu$, as well as their proposed HL-LHC successors FASER2 and FASER$\nu$2. 
We considered a class of scenarios in which the HNLs couple directly to a new $Z'$ gauge boson, while their visible decays arise through mixing with the SM neutrinos. 
As representative examples, we studied the $U(1)_{B-L}$, $U(1)_{B-3L_\tau}$, and $U(1)_B$ gauge symmetries, together with both flavor-democratic and $\tau$-dominated HNL mixing benchmarks.

A central feature of this setup is that unlike the canonical case, the production of HNLs is not controlled solely by their small active-sterile mixing. 
Instead, HNL pairs can be produced efficiently through an intermediate mediator $Z'$, allowing the production rate and the HNL decay probability to be governed by parametrically different interactions. 
We considered both proton bremsstrahlung (as formulated in Section~\ref{sec:brem}) and Drell-Yan production (as in Section~\ref{sec:DY}) of $Z'$. 
These two mechanisms probe complementary mediator-mass regimes: bremsstrahlung is most relevant for relatively light mediators ($m_{Z'}<1$ GeV), whereas Drell-Yan production becomes increasingly important at larger masses, as shown in Fig.~\ref{Fig.HNL_eff_and_Flux}(a). 
Combining the two mechanisms therefore provides sensitivity over a broad range of HNL and mediator masses.

We have examined several possible production locations relevant to the far-forward LHC environment, as listed in Section~\ref{sec:Exp_Settings}. 
In addition to the dominant production at the ATLAS IP, we studied HNL production initiated by the large flux of forward neutrons impinging on the TAN/TAXN absorbers, as well as by secondary protons, neutrons, and charged pions produced in these absorbers. 
Although these secondary interactions provide an additional conceptual production mechanism, our simulations indicate that their contribution to the observable HNL event rate is subleading compared with production at the ATLAS IP for the scenarios considered here. The far-forward geometry of the signal, together with the increased luminosity and enlarged detector volumes envisioned for the HL-LHC, makes FASER2 and FASER$\nu$2 particularly powerful probes of this class of models.

Using the four experiments and three benchmark models mentioned above, we considered visible HNL decays in the $N\to\nu\pi^0$, $N\to\nu\ell^+\ell^-$, and $N\to\ell^\pm+\text{hadrons}$ channels. For each experiment, we accounted for the geometric acceptance, the probability that the HNL decays within the detector, the energy requirements, and representative particle-identification efficiencies. We used two flavor-mixing benchmarks: 111, in which the HNL mixes equally with all three neutrino flavors, and 001, in which it mixes only with $\nu_\tau$. Unless stated otherwise, the sensitivities to $|U|$ quoted below are for the $U(1)_B$ model and apply to both flavor benchmarks.

We first presented the $90\%$ C.L.~sensitivities in the $|U|^2$--$m_N$ plane for the $N\to\nu\pi^0$, $N\to\nu e^+e^-$, and $N\to\nu\mu^+\mu^-$ channels in Fig.~\ref{Fig.HNL_fluxnuXevents}, with panels (a)--(c) for the 111 benchmark and (d)--(f) for the 001 benchmark. For bremsstrahlung production, the $\rho$ and $\omega$ resonances cause a sharp rise in sensitivity at $m_N\approx 0.4$~GeV. In particular, FASER2 reaches $|U|\gtrsim 10^{-4}$ in the $\nu\pi^0$ channel, while for the 001 benchmark, FASER (in $\nu\pi^0$), FASER2 (in $\nu\ell^+\ell^-$), and FASER$\nu$2 (in $\nu\ell^+\ell^-$) each reach $|U|\gtrsim 10^{-3}$.
Drell-Yan production extends the reach to multi-GeV masses. FASER has similar sensitivity in both the $\nu\pi^0$ and $\nu\ell^+\ell^-$ channels, reaching $|U|\gtrsim 10^{-3}$ at $m_N\approx 3$~GeV. At FASER$\nu$ and FASER$\nu$2, muons are identified more efficiently than electrons, which favors the $\nu\mu^+\mu^-$ channel. It is the most sensitive channel at FASER$\nu$, reaching $|U|\gtrsim 3\times10^{-3}$ at $m_N\approx 4$~GeV, and at FASER$\nu$2 it reaches $|U|\gtrsim 10^{-4}$ at $m_N\approx 6$~GeV. FASER2 does significantly better in the $\nu\ell^+\ell^-$ channels, reaching $|U|\gtrsim 10^{-5}$ at $m_N\approx 6$~GeV, thanks to the similar $\nu e^+e^-$ and $\nu\mu^+\mu^-$ branching ratios together with its higher luminosity and larger decay volume. In the $U(1)_{B-3L_\tau}$ model, the corresponding reaches are $|U|\gtrsim 3\times10^{-3}$ at $m_N\approx 3$~GeV for FASER$\nu$2 and $|U|\gtrsim 10^{-4}$ at $m_N\approx 18$~GeV for FASER2.

We then turned to the $N\to\ell^\pm+\text{hadrons}$ channels in Fig.~\ref{Fig.HNL_fluxlhadronsevents}, with $\ell=e$, $\mu$, and $\tau$ shown in panels (a)--(c) for the 111 benchmark and $\ell=\tau$ (d) for the 001 benchmark. Bremsstrahlung production contributes less here, since at the lower masses it accesses, $N\to\ell^\pm\pi^\mp$ is subdominant to $N\to\nu\pi^0$. By contrast, because this is the dominant HNL decay mode for $m_N\gtrsim 1$~GeV, Drell-Yan produced HNLs decaying to these channels achieve substantial sensitivities. FASER and FASER$\nu$ both reach $|U|\gtrsim 10^{-3}$ at $m_N\approx 5$~GeV, in the $e$ and $\mu$ channels for the 111 benchmark and in the $\tau$ channel for the 001 benchmark. FASER2 and FASER$\nu$2 extend well beyond the excluded parameter space for all three lepton flavors: FASER$\nu$2 reaches $|U|\gtrsim 10^{-5}$ at $m_N\approx 10$~GeV, and FASER2 reaches $|U|\gtrsim 3\times10^{-6}$ at $m_N\approx 15$~GeV and $|U|\gtrsim 3\times10^{-5}$ at $m_N\approx 8$~GeV in the $U(1)_{B-3L_\tau}$ model. Impressively, for our $U(1)_B$ model benchmark, FASER2 reaches the Type-I seesaw expectation for all three lepton flavors in the 111 benchmark, as well as for the $\tau$ flavor in the 001 benchmark.

Finally, we presented the sensitivity to the gauge coupling $g_X$ in the $U(1)_{B-3L_\tau}$ and $U(1)_B$ models, using the $N\to\nu\pi^0$ and $N\to\nu\ell^+\ell^-$ channels in Fig.~\ref{fig:gauge_nuX} and the $N\to\ell^\pm+\text{hadrons}$ channels in Fig.~\ref{fig:lhadrons_gaugeSensitivities}. We fixed $m_{Z'}/m_N = 2.1$ and took $|U| = 3\times10^{-4}$ for the 111 benchmark and $|U| = 10^{-3}$ for the 001 benchmark. FASER can probe couplings down to $g_X\approx 10^{-2}$ at $m_N\approx 6$~GeV in the $\nu\pi^0$ and $\ell^\pm+\text{hadrons}$ channels, and FASER$\nu$ reaches a similar value in the same channels apart from $\nu\pi^0$. FASER2 reaches $g_X\approx 2\times10^{-4}$ at $m_N\approx 5$~GeV in the $\nu\ell^+\ell^-$ and $\ell^\pm+\text{hadrons}$ channels, and FASER$\nu$2 reaches $g_X\approx 2\times10^{-3}$ at $m_N\approx 6$~GeV in the $\nu\mu^+\mu^-$ and $\ell^\pm+\text{hadrons}$ channels. Overall, FASER2 and FASER$\nu$2 can probe beyond the excluded $U(1)_{B-3L_\tau}$ parameter space, and all four experiments can do so for $U(1)_B$.

Our results show that the combination of mediator-assisted production and long-lived HNL decays enable the FPF program to probe regions of the HNL mass-mixing parameter space beyond existing constraints, including regions approaching the mixing strengths motivated by low-scale neutrino mass generation mechanisms. 
The different decay channels and detector concepts provide complementary sensitivity, particularly between the conventional FASER detectors and the emulsion-based FASER$\nu$ detectors.

More broadly, our results illustrate that the far-forward LHC program can probe HNL scenarios that differ qualitatively from the conventional picture in which both production and decay are controlled by active-sterile mixing. 
Additional interactions responsible for HNL production can substantially enhance their flux while leaving the long lifetime required for displaced decays intact. 
This makes FASER, FASER$\nu$, and especially their proposed HL-LHC successors sensitive probes of a broad class of gauge-extended HNL models and provides further motivation for exploiting the far-forward region of the LHC in searches for new weakly coupled states.

\section*{Acknowledgments}
We thank Wolfgang Kilian for implementing $\pi p$ scattering in \texttt{WHIZARD} for our work and Hyunyong Kim for his work on the \texttt{GEANT4} simulations of secondary particles used here. We also thank  Felix Kling and Adam Ritz for helpful discussions. FB and TH were supported in part by the U.S.~Department of Energy under grant No.~DE-SC0007914 and in  part by Pitt PACC. PSBD was supported in part by the US Department of Energy under grant No. DE-SC0017987. BD and AK are supported in part by the U.S. Department of Energy Grant~DE-SC0010813. The work of DK is supported in part by NSF Grant No. PHY-2609913. This work was produced by Fermi Forward Discovery Group, LLC under Contract No. 89243024CSC000002 with the U.S. Department of Energy, Office of Science, Office of High Energy Physics. Publisher acknowledges the U.S. Government license to provide public access under the \href{https://www.energy.gov/doe-public-access-plan}{DOE Public Access Plan}.

\FloatBarrier

\appendix
\FloatBarrier

\section{Bremsstrahlung Production}
\label{app:brem}

In the QR-ISR approximation, the matrix element for the bremsstrahlung of a vector boson $V$ from a nucleon with four-momentum $p_N$ as it interacts with a target with four-momentum $p_t$, $\mathcal{M}^{p_N p_t \to V f}$, can be expressed in a form that is proportional to the general matrix element for diffractive processes, $\mathcal{M}^{p_N p_t \to f}$. The resultant differential cross section for ISR bremsstrahlung is given by Eq.~\eqref{eq:bremcs}, which is written in terms of a splitting function. This is the sum of contributions arising from the vector-like current that appears in the proton-proton-$V$ vertex. 

Given that the gauge boson corresponds to a symmetry $U(1)_X$, the current can be written using the Gordon decomposition:  
\begin{equation}
    \langle N| J^\mu_X|N\rangle = \bar{u}(p')\left[ \gamma^\mu F_{1,X}^{N}(t)+\frac{i\sigma^{\mu\nu}(p'-p)_\nu}{2m_N} F_{2,X}^N(t)\right] u(p) , 
    \label{eq:current}
\end{equation}
where $F_{1,X}^{N} (t)$ and $F_{2,X}^{N} (t)$ are the Dirac and Pauli form factors of the nucleon $N$, respectively, that satisfy the boundary conditions determined by the nucleon's charge under the $U(1)_X$ symmetry. The variable $t = (p'-p)^2$ is the four-momentum square transferred to $V$. Since we consider the on-shell production of gauge bosons, we will set $t = k^2 = m_V^2$. The current~\eqref{eq:current} appears while expanding the matrix-element square. Therefore, the total splitting function can be written in terms of individual Dirac and Pauli splitting functions $\omega_1$ and $\omega_2$ as well as the corresponding form factors, as follows:
\begin{equation}
    \begin{aligned}
        \omega_N(z, p_T^2) = & \big[ \omega_1(z,p_T^2)|F_{1,X}^{N}(k^2)|^2+\omega_2(z,p_T^2)|F_{2,X}^{N}(k^2)|^2 \\
        &+ \omega_{12}(z,p_T^2)\text{Re}(F_{1,X}^{N}(k^2)F_{2,X}^{*N}(k^2))\big]\kappa^2_{pp'V}(p'^2) ,
    \end{aligned}
\end{equation}
where $\kappa_{pp'V}(p'^2)$ is a time-like form factor that imposes the condition that the intermediate nucleon with four-momentum $p'$ is close to being on-shell. This is imposed through a dipole form factor
\begin{equation}
    \kappa_{pp'V}(p'^2) = \frac{1}{1 + (p'^2-m_p^2)^2/\Lambda_p^4} \, ,
\end{equation}
where we fix $\Lambda_p = 1.5$~GeV in this analysis.
The nucleon form factors can be written as a sum over the Breit-Wigner forms corresponding to the isovector and isoscalar vector-like mesons $\rho$, $\omega$, and $\phi$, by including the appropriate charges of the mesons under the $U(1)_X$ symmetry~\cite{Kling:2025udr}. Furthermore, this formulation is valid only when the intermediate proton in the scattering process is very close to being on-shell, and when the outgoing vector boson is high in energy and forward. These conditions can be encompassed into the following relations:  
\begin{equation}
    \frac{H}{4z(1-z)p_n^2} \ll 1, \qquad p_T, m_p,m_V \ll E_p,E_V\ ,
\end{equation}
where $H$ is defined as $H = z(m_p^2 - (p_N-k)^2)$.

\section{HNLs from Primary Neutrons and Secondary Hadrons}
\label{App:Secondary_Fluxes}

In fixed-target or beam-dump experiments, one relies on the large number of incoming protons to hit the target, which is often measured by POT.  In our study, we consider not just protons hitting the target, but also other hadrons on target (HOT) such as neutrons and pions. We henceforth denote the incoming hadron as $h$, and we adapt the existing treatment for protons (as shown in e.g., Ref.~\cite{Batell:2020vqn}) to HOTs. The predicted BSM signal rate under consideration depends on the production cross section of a mediator particle 
$X$, $\sigma(hp\to X)$  and its decay branching ratio to the observable SM final state ${\rm BR}(X\to {\rm final\ state})$ with detection efficiency $\text{eff}_f$ as defined in Eq.~\eqref{eq:eff}. 
We adopt the total cross section per nucleon $\sigma_{hp}(s)$ as parameterized in Ref.~\cite{Block:2005pt} for pions and protons. We account for the material effects of the iron target by noting the scaling $\sigma_{hA}(s) \approx A^{\alpha_h}\sigma_{hp}(s)$, where $A=56$ is the mass number of the iron nucleus and we take $\alpha_{\pi^{\pm}} \approx 0.75$ for pions and $\alpha_p 
\approx 0.7$ for protons~\cite{Denisov:1973zv}. Noting that for Drell-Yan production $\sigma_{hA} \approx A\sigma_{hN}$, we arrive at the following expression for fixed targets:  
\begin{equation}
N_{\text{Detected}} = \text{HOT}\times A^{1 - \alpha_{h}}\frac{\sigma(hp\to X)}{\sigma_{hp}(s)}\times {\rm BR}(X\to {\rm final\  state})\times\text{eff}_f.
\label{eq:events} 
\end{equation}

\begin{figure}[ht!]
    \centering 
\includegraphics[width=0.32\linewidth]{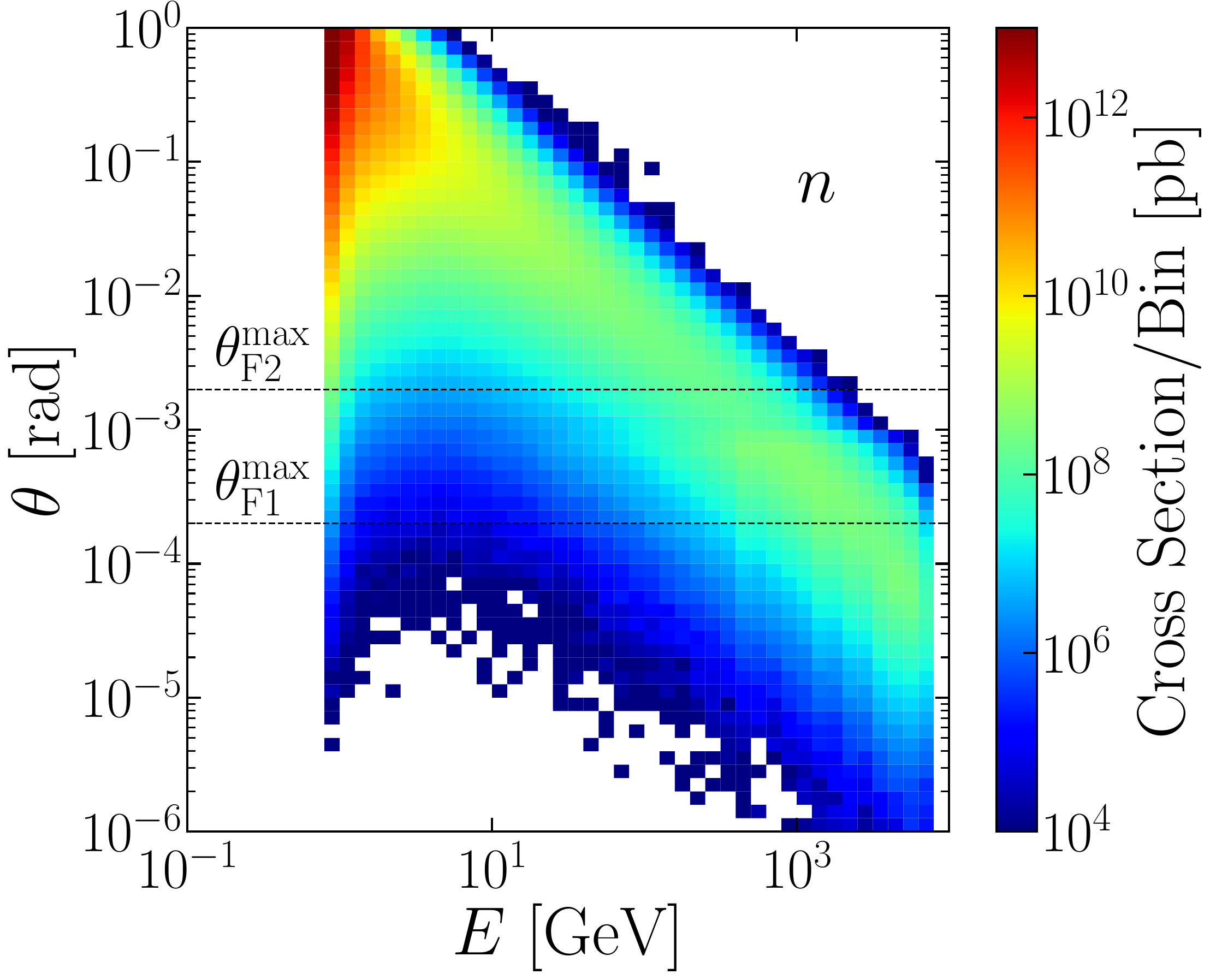}
\includegraphics[width=0.32\linewidth]{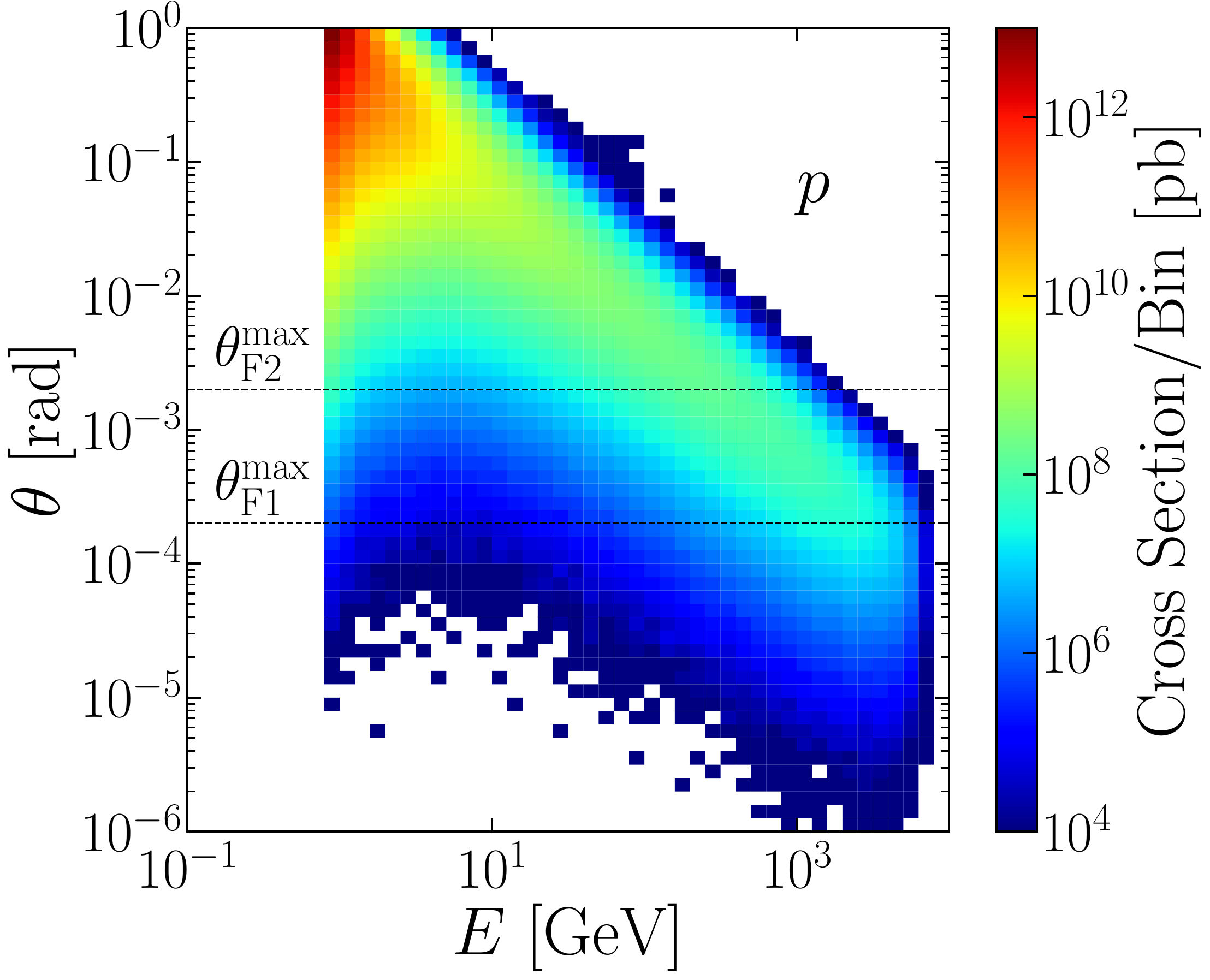}
\includegraphics[width=0.32\linewidth]{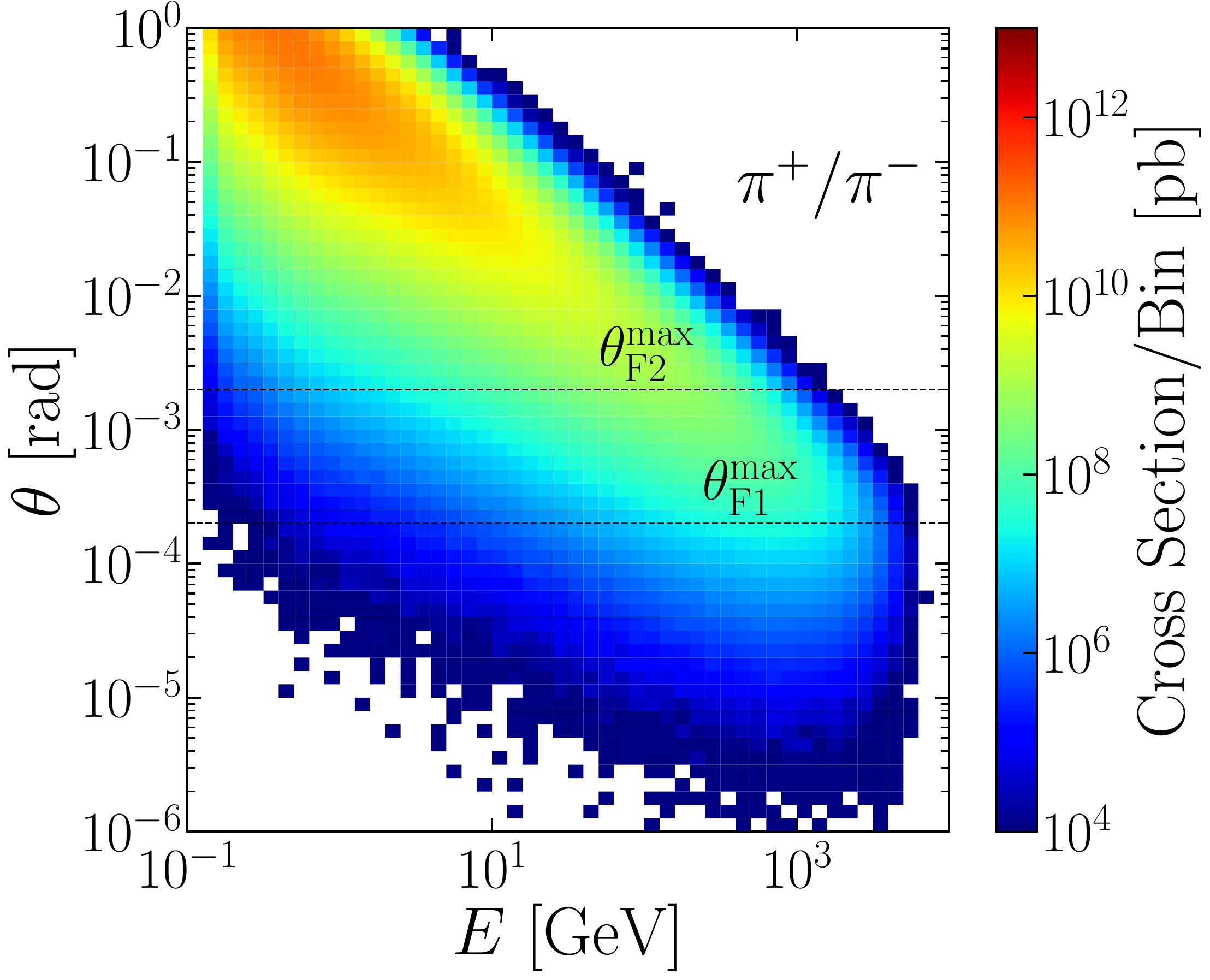}
   \caption{ Hadron cross sections at the iron absorber TAN/TAXN.
}
\label{Fig.Hadron_Fluxes}
\end{figure}

We used \texttt{GEANT4}~\cite{GEANT4:2002zbu} to obtain our distributions of neutron primaries and other hadron secondaries produced within an iron target. In Fig.~\ref{Fig.Hadron_Fluxes}, we present the cross sections from neutron primaries and secondaries, proton secondaries, and $\pi^+$, $\pi^-$ secondaries. In each panel, we denote the maximum angular acceptances for FASER and FASER2 by $\theta_{\text{F1}}^{\text{max}}\approx 0.0002$ and $\theta_{\text{F1}}^{\text{max}}\approx 0.002$, respectively (cf.~Eq.~\eqref{eq:angle}). We find that the majority of the secondaries are produced at small energy and high angles. The neutron primaries, with mostly high energies and very small angles, thus provide the larger contribution to our HNL production. 

\begin{figure}[tb]
    \centering 
\includegraphics[width=0.495\linewidth]{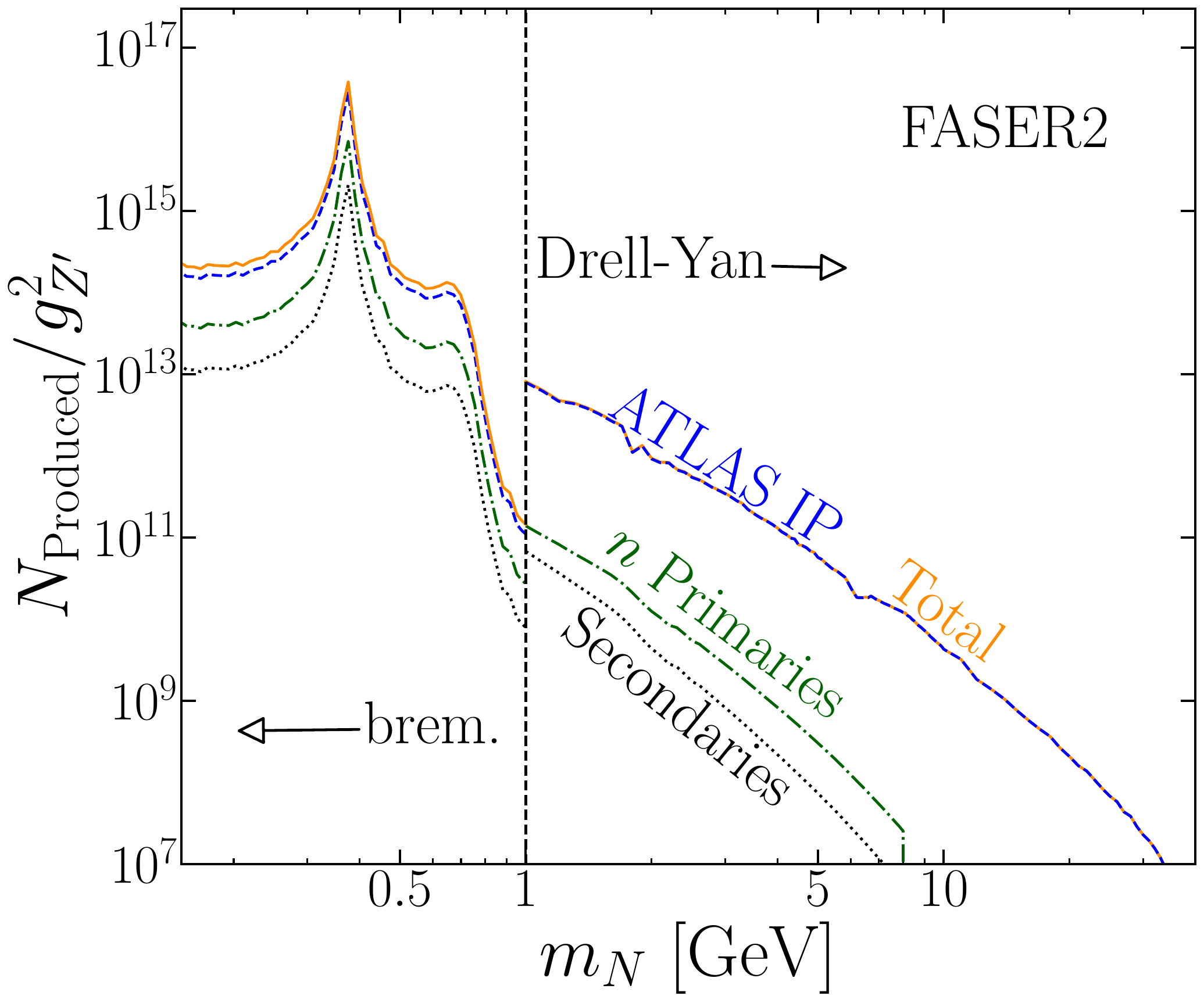}
   \caption{HNL flux within the geometric acceptance of FASER2 showing the individual contributions from ATLAS IP (blue dashed), neutron primaries (green dot-dashed), secondary particles (black dotted) and the total flux (orange solid) from bremsstrahlung and Drell-Yan production modes. 
}
\label{Fig.HNLflux_nevents}
\end{figure}

In Fig.~\ref{Fig.HNLflux_nevents}, we present each channel's contribution to the HNL flux at FASER2 as a representative example. Across both bremsstrahlung and Drell-Yan modes, we observe that the HNLs arising from the ATLAS IP production constitute the majority of events, with primary neutrons and the secondary production modes providing a sub-leading contribution.

\bibliographystyle{JHEP}
\bibliography{references}

@article{Atre:2009rg,
    author = "Atre, Anupama and Han, Tao and Pascoli, Silvia and Zhang, Bin",
    title = "{The Search for Heavy Majorana Neutrinos}",
    eprint = "0901.3589",
    archivePrefix = "arXiv",
    primaryClass = "hep-ph",
    reportNumber = "FERMILAB-PUB-08-086-T, NSF-KITP-08-54, MADPH-06-1466, DCPT-07-198, IPPP-07-99",
    doi = "10.1088/1126-6708/2009/05/030",
    journal = "JHEP",
    volume = "05",
    pages = "030",
    year = "2009"
}

@article{Kilian:2007gr,
    author = "Kilian, Wolfgang and Ohl, Thorsten and Reuter, Jurgen",
    title = "{WHIZARD: Simulating Multi-Particle Processes at LHC and ILC}",
    eprint = "0708.4233",
    archivePrefix = "arXiv",
    primaryClass = "hep-ph",
    reportNumber = "DESY-11-126, EDINBURGH-2010-36, FR-PHENO-2010-037, SI-HEP-2010-18",
    doi = "10.1140/epjc/s10052-011-1742-y",
    journal = "Eur. Phys. J. C",
    volume = "71",
    pages = "1742",
    year = "2011"
}

@article{Burk:2026mvk,
    author = "Burk, Francis M. and Dev, P. S. Bhupal and Dutta, Bhaskar and Han, Tao and Karthikeyan, Aparajitha and Kim, Doojin",
    title = "{Drell-Yan Production of New Particles at Fixed-Target Experiments: Heavy Neutral Lepton as a Case Study}",
    eprint = "2601.18874",
    archivePrefix = "arXiv",
    primaryClass = "hep-ph",
    reportNumber = "PITT-PACC-2510, MI-HET-877",
    month = "1",
    year = "2026"
}

@article{Feng:2022inv,
    author = "Feng, Jonathan L. and others",
    title = "{The Forward Physics Facility at the High-Luminosity LHC}",
    eprint = "2203.05090",
    archivePrefix = "arXiv",
    primaryClass = "hep-ex",
    reportNumber = "UCI-TR-2022-01, CERN-PBC-Notes-2022-001, INT-PUB-22-006, BONN-TH-2022-04, FERMILAB-PUB-22-094-ND-SCD-T",
    doi = "10.1088/1361-6471/ac865e",
    journal = "J. Phys. G",
    volume = "50",
    number = "3",
    pages = "030501",
    year = "2023"
}

@article{Tastet:2021vwp,
    author = "Tastet, Jean-Loup and Ruchayskiy, Oleg and Timiryasov, Inar",
    title = "{Reinterpreting the ATLAS bounds on heavy neutral leptons in a realistic neutrino oscillation model}",
    eprint = "2107.12980",
    archivePrefix = "arXiv",
    primaryClass = "hep-ph",
    doi = "10.1007/JHEP12(2021)182",
    journal = "JHEP",
    volume = "12",
    pages = "182",
    year = "2021"
}

@article{Abada:2022wvh,
    author = "Abada, Asmaa and Escribano, Pablo and Marcano, Xabier and Piazza, Gioacchino",
    title = "{Collider searches for heavy neutral leptons: beyond simplified scenarios}",
    eprint = "2208.13882",
    archivePrefix = "arXiv",
    primaryClass = "hep-ph",
    reportNumber = "IFT-UAM/CSIC-22-98",
    doi = "10.1140/epjc/s10052-022-11011-7",
    journal = "Eur. Phys. J. C",
    volume = "82",
    number = "11",
    pages = "1030",
    year = "2022"
}

@article{FASER:2023tle,
    author = "Abreu, Henso and others",
    collaboration = "FASER",
    title = "{Search for dark photons with the FASER detector at the LHC}",
    eprint = "2308.05587",
    archivePrefix = "arXiv",
    primaryClass = "hep-ex",
    reportNumber = "CERN-EP-2023-161",
    doi = "10.1016/j.physletb.2023.138378",
    journal = "Phys. Lett. B",
    volume = "848",
    pages = "138378",
    year = "2024"
}

@article{FASER:2018eoc,
    author = "Ariga, Akitaka and others",
    collaboration = "FASER",
    title = "{FASER{\textquoteright}s physics reach for long-lived particles}",
    eprint = "1811.12522",
    archivePrefix = "arXiv",
    primaryClass = "hep-ph",
    reportNumber = "UCI-TR-2018-19, KYUSHU-RCAPP-2018-06",
    doi = "10.1103/PhysRevD.99.095011",
    journal = "Phys. Rev. D",
    volume = "99",
    number = "9",
    pages = "095011",
    year = "2019"
}

@article{FASER:2026jhh,
    author = "Mammen Abraham, Roshan and others",
    collaboration = "FASER",
    title = "{Electromagnetic Shower Reconstruction and Identification in FASER's Emulsion Detector for LHC Forward Neutrino Measurements}",
    eprint = "2606.18517",
    archivePrefix = "arXiv",
    primaryClass = "hep-ex",
    reportNumber = "CERN-FASER-2026-001",
    month = "6",
    year = "2026"
}

@article{Kling:2023tgr,
    author = {Kling, Felix and M{\"a}kel{\"a}, Toni and Trojanowski, Sebastian},
    title = "{Investigating the fluxes and physics potential of LHC neutrino experiments}",
    eprint = "2309.10417",
    archivePrefix = "arXiv",
    primaryClass = "hep-ph",
    reportNumber = "DESY-23-131",
    doi = "10.1103/PhysRevD.108.095020",
    journal = "Phys. Rev. D",
    volume = "108",
    number = "9",
    pages = "095020",
    year = "2023"
}

@article{FASER:2022hcn,
    author = "Abreu, Henso and others",
    collaboration = "FASER",
    title = "{The FASER detector}",
    eprint = "2207.11427",
    archivePrefix = "arXiv",
    primaryClass = "physics.ins-det",
    reportNumber = "CERN-FASER-2022-001",
    doi = "10.1088/1748-0221/19/05/P05066",
    journal = "JINST",
    volume = "19",
    number = "05",
    pages = "P05066",
    year = "2024"
}

@article{T2K:2019jwa,
    author = "Abe, K. and others",
    collaboration = "T2K",
    title = "{Search for heavy neutrinos with the T2K near detector ND280}",
    eprint = "1902.07598",
    archivePrefix = "arXiv",
    primaryClass = "hep-ex",
    doi = "10.1103/PhysRevD.100.052006",
    journal = "Phys. Rev. D",
    volume = "100",
    number = "5",
    pages = "052006",
    year = "2019"
}

@article{CMS:2024ita,
    author = "Hayrapetyan, Aram and others",
    collaboration = "CMS",
    title = "{Search for long-lived heavy neutrinos in the decays of B mesons produced in proton-proton collisions at $ \sqrt{s} $ = 13 TeV}",
    eprint = "2403.04584",
    archivePrefix = "arXiv",
    primaryClass = "hep-ex",
    reportNumber = "CMS-EXO-22-019, CERN-EP-2024-053",
    doi = "10.1007/JHEP06(2024)183",
    journal = "JHEP",
    volume = "06",
    pages = "183",
    year = "2024"
}

@article{ATLAS:2025uah,
    author = "Aad, Georges and others",
    collaboration = "ATLAS",
    title = "{Search for heavy neutral leptons in decays of W bosons using leptonic and semi-leptonic displaced vertices in $ \sqrt{s} $ = 13 TeV pp collisions with the ATLAS detector}",
    eprint = "2503.16213",
    archivePrefix = "arXiv",
    primaryClass = "hep-ex",
    reportNumber = "CERN-EP-2025-052",
    doi = "10.1007/JHEP07(2025)196",
    journal = "JHEP",
    volume = "07",
    pages = "196",
    year = "2025"
}

@article{ATLAS:2024fcs,
    author = "Aad, Georges and others",
    collaboration = "ATLAS",
    title = "{Search for heavy right-handed Majorana neutrinos in the decay of top quarks produced in proton-proton collisions at s=13{\,}{\,}TeV with the ATLAS detector}",
    eprint = "2408.05000",
    archivePrefix = "arXiv",
    primaryClass = "hep-ex",
    reportNumber = "CERN-EP-2024-154",
    doi = "10.1103/PhysRevD.110.112004",
    journal = "Phys. Rev. D",
    volume = "110",
    number = "11",
    pages = "112004",
    year = "2024"
}

@article{ATLAS:2025lva,
    author = "Aad, Georges and others",
    collaboration = "ATLAS",
    title = "{Search for heavy neutral leptons in decays of W bosons produced in 13 TeV $pp$ collisions using prompt signatures in the ATLAS detector}",
    eprint = "2508.20929",
    archivePrefix = "arXiv",
    primaryClass = "hep-ex",
    reportNumber = "CERN-EP-2025-164",
    doi = "10.1140/epjc/s10052-025-15191-w",
    journal = "Eur. Phys. J. C",
    volume = "86",
    number = "2",
    pages = "153",
    year = "2026"
}

@article{CMS:2022fut,
    author = "Tumasyan, Armen and others",
    collaboration = "CMS",
    title = "{Search for long-lived heavy neutral leptons with displaced vertices in proton-proton collisions at $ \sqrt{\mathrm{s}} $ =13 TeV}",
    eprint = "2201.05578",
    archivePrefix = "arXiv",
    primaryClass = "hep-ex",
    reportNumber = "CMS-EXO-20-009, CERN-EP-2021-264",
    doi = "10.1007/JHEP07(2022)081",
    journal = "JHEP",
    volume = "07",
    pages = "081",
    year = "2022"
}

@article{CMS:2024xdq,
    author = "Hayrapetyan, Aram and others",
    collaboration = "CMS",
    title = "{Search for heavy neutral leptons in final states with electrons, muons, and hadronically decaying tau leptons in proton-proton collisions at $ \sqrt{s} $ = 13 TeV}",
    eprint = "2403.00100",
    archivePrefix = "arXiv",
    primaryClass = "hep-ex",
    reportNumber = "CMS-EXO-22-011, CERN-EP-2024-032",
    doi = "10.1007/JHEP06(2024)123",
    journal = "JHEP",
    volume = "06",
    pages = "123",
    year = "2024"
}

@article{CMS:2018iaf,
    author = "Sirunyan, Albert M and others",
    collaboration = "CMS",
    title = "{Search for heavy neutral leptons in events with three charged leptons in proton-proton collisions at $\sqrt{s} =$ 13 TeV}",
    eprint = "1802.02965",
    archivePrefix = "arXiv",
    primaryClass = "hep-ex",
    reportNumber = "CMS-EXO-17-012, CERN-EP-2018-006",
    doi = "10.1103/PhysRevLett.120.221801",
    journal = "Phys. Rev. Lett.",
    volume = "120",
    number = "22",
    pages = "221801",
    year = "2018"
}

@article{ATLAS:2022atq,
    author = "Aad, Georges and others",
    collaboration = "ATLAS",
    title = "{Search for Heavy Neutral Leptons in Decays of W Bosons Using a Dilepton Displaced Vertex in s=13{\,}{\,}TeV pp Collisions with the ATLAS Detector}",
    eprint = "2204.11988",
    archivePrefix = "arXiv",
    primaryClass = "hep-ex",
    reportNumber = "CERN-EP-2022-017",
    doi = "10.1103/PhysRevLett.131.061803",
    journal = "Phys. Rev. Lett.",
    volume = "131",
    number = "6",
    pages = "061803",
    year = "2023"
}

@article{CMS:2024ake,
    author = "Hayrapetyan, Aram and others",
    collaboration = "CMS",
    title = "{Search for long-lived heavy neutral leptons decaying in the CMS muon detectors in proton-proton collisions at s=13{\,}{\,}TeV}",
    eprint = "2402.18658",
    archivePrefix = "arXiv",
    primaryClass = "hep-ex",
    reportNumber = "CMS-EXO-22-017, CERN-EP-2024-022",
    doi = "10.1103/PhysRevD.110.012004",
    journal = "Phys. Rev. D",
    volume = "110",
    pages = "012004",
    year = "2024"
}

@article{FASER:2019dxq,
    author = "Abreu, Henso and others",
    collaboration = "FASER",
    title = "{Detecting and Studying High-Energy Collider Neutrinos with FASER at the LHC}",
    eprint = "1908.02310",
    archivePrefix = "arXiv",
    primaryClass = "hep-ex",
    reportNumber = "CERN-EP-2019-160, KYUSHU-RCAPP-2019-003, SLAC-PUB-17460, UCI-TR-2019-19, SLAC-PUB-17460,
  UCI-TR-2019-19",
    doi = "10.1140/epjc/s10052-020-7631-5",
    journal = "Eur. Phys. J. C",
    volume = "80",
    number = "1",
    pages = "61",
    year = "2020"
}

@misc{kose,
  author = {Umut Kose},
  title  = "{TeV-Scale Neutrino Physics with FASER
at the LHC}",
  howpublished = {\url{https://indico.cern.ch/event/1639855/contributions/7194037/attachments/3357069/6021361/DarkInt_UKose_FASER_Neutrino.pdf}}
}

@article{Arguelles:2021dqn,
    author = {Arg{\"u}elles, Carlos A. and Foppiani, Nicol{\`o} and Hostert, Matheus},
    title = "{Heavy neutral leptons below the kaon mass at hodoscopic neutrino detectors}",
    eprint = "2109.03831",
    archivePrefix = "arXiv",
    primaryClass = "hep-ph",
    doi = "10.1103/PhysRevD.105.095006",
    journal = "Phys. Rev. D",
    volume = "105",
    number = "9",
    pages = "095006",
    year = "2022"
}

@article{Blumlein:2013cua,
    author = {Bl{\"u}mlein, Johannes and Brunner, J{\"u}rgen},
    title = "{New Exclusion Limits on Dark Gauge Forces from Proton Bremsstrahlung in Beam-Dump Data}",
    eprint = "1311.3870",
    archivePrefix = "arXiv",
    primaryClass = "hep-ph",
    reportNumber = "DESY-13-202, DO-TH-13-29, SFB-CPP-13-87, LPN-13-087",
    doi = "10.1016/j.physletb.2014.02.029",
    journal = "Phys. Lett. B",
    volume = "731",
    pages = "320--326",
    year = "2014"
}

@article{FPF:2025bor,
    author = "Anchordoqui, Luis A. and others",
    collaboration = "FPF",
    title = "{The forward physics facility: Physics opportunities and conceptual design}",
    eprint = "2510.26260",
    archivePrefix = "arXiv",
    primaryClass = "hep-ex",
    reportNumber = "CERN-PBC-Notes-2025-010",
    doi = "10.1016/j.nuclphysb.2026.117398",
    journal = "Nucl. Phys. B",
    volume = "1026",
    pages = "117398",
    year = "2026"
}

@article{vbbq-fb1t,
  title = {Characterizing heavy neutral leptons: Measuring parameters, discriminating Majorana versus Dirac, and using FASER2 as a trigger for ATLAS},
  author = {Feng, Jonathan L. and Hewitt, Alec and La Rocco, Daniel  and Whiteson, Daniel},
  journal = {Phys. Rev. D},
  volume = {113},
  issue = {3},
  pages = {035006},
  numpages = {15},
  year = {2026},
  month = {Feb},
  publisher = {American Physical Society},
  doi = {10.1103/vbbq-fb1t},
  url = {https://link.aps.org/doi/10.1103/vbbq-fb1t}
}

@inproceedings{FPFWorkingGroups:2025rsc,
    author = "Anchordoqui, Luis A. and others",
    collaboration = "FPF Working Groups",
    title = "{The Forward Physics Facility at the Large Hadron Collider}",
    eprint = "2503.19010",
    archivePrefix = "arXiv",
    primaryClass = "hep-ex",
    month = "3",
    year = "2025"
}

@article{MammenAbraham:2025gai,
    author = "Mammen Abraham, Roshan and Fieg, Max",
    title = "{LHC as a TeV muon beam dump: Muonphilic scalars at FASER}",
    eprint = "2501.09071",
    archivePrefix = "arXiv",
    primaryClass = "hep-ph",
    doi = "10.1103/11j5-txpc",
    journal = "Phys. Rev. D",
    volume = "112",
    number = "1",
    pages = "015005",
    year = "2025"
}

@techreport{Boyd:2803084,
      author        = "Boyd, Jamie",
      title         = "The FASER W-Si High Precision Preshower  Technical Proposal",
      institution   = "CERN",
      reportNumber  = "CERN-LHCC-2022-006, LHCC-P-023",
      address       = "Geneva",
      year          = "2022",
      number="",
      url           = "https://cds.cern.ch/record/2803084"
}

@article{FASER:2018bac,
    author = "Ariga, Akitaka and others",
    collaboration = "FASER",
    title = "{Technical Proposal for FASER: ForwArd Search ExpeRiment at the LHC}",
    eprint = "1812.09139",
    archivePrefix = "arXiv",
    primaryClass = "physics.ins-det",
    reportNumber = "CERN-LHCC-2018-036, LHCC-P-013, UCI-TR-2018-22, KYUSHU-RCAPP-2018-07",
    month = "12",
    year = "2018"
}

@article{FASER:2020gpr,
    author = "Abreu, Henso and others",
    collaboration = "FASER",
    title = "{Technical Proposal: FASERnu}",
    eprint = "2001.03073",
    archivePrefix = "arXiv",
    primaryClass = "physics.ins-det",
    reportNumber = "CERN-LHCC-2019-017, LHCC-P-015, UCI-TR-2019-25",
    month = "1",
    year = "2020"
}

@article{Denisov:1973zv,
    author = "Denisov, S. P. and Donskov, S. V. and Gorin, Yu. P. and Krasnokutsky, R. N. and Petrukhin, A. I. and Prokoshkin, Yu. D. and Stoyanova, D. A.",
    title = "{Absorption cross-sections for pions, kaons, protons and anti-protons on complex nuclei in the 6-GeV/c to 60-GeV/c momentum range}",
    doi = "10.1016/0550-3213(73)90351-9",
    journal = "Nucl. Phys. B",
    volume = "61",
    pages = "62--76",
    year = "1973"
}

@article{Block:2005pt,
    author = "Block, M. M. and Halzen, F.",
    title = "{New evidence for the saturation of the Froissart bound}",
    eprint = "hep-ph/0506031",
    archivePrefix = "arXiv",
    reportNumber = "NUHEP-989, UW-MADPH-04-1382",
    doi = "10.1103/PhysRevD.72.039902",
    journal = "Phys. Rev. D",
    volume = "72",
    pages = "036006",
    year = "2005",
    note = "[Erratum: Phys.Rev.D 72, 039902 (2005)]"
}

@article{DELPHI:1996qcc,
    author = "Abreu, P. and others",
    collaboration = "DELPHI",
    title = "{Search for neutral heavy leptons produced in Z decays}",
    reportNumber = "CERN-PPE-96-195",
    doi = "10.1007/s002880050370",
    journal = "Z. Phys. C",
    volume = "74",
    pages = "57--71",
    year = "1997",
    note = "[Erratum: Z.Phys.C 75, 580 (1997)]"
}

@article{Abdullahi:2022jlv,
    author = "Abdullahi, Asli M. and others",
    title = "{The present and future status of heavy neutral leptons}",
    eprint = "2203.08039",
    archivePrefix = "arXiv",
    primaryClass = "hep-ph",
    reportNumber = "FERMILAB-CONF-22-184-T-V",
    doi = "10.1088/1361-6471/ac98f9",
    journal = "J. Phys. G",
    volume = "50",
    number = "2",
    pages = "020501",
    year = "2023"
}

@article{MicroBooNE:2015bmn,
    author = "Acciarri, R. and others",
    collaboration = "MicroBooNE, LAr1-ND, ICARUS-WA104",
    title = "{A Proposal for a Three Detector Short-Baseline Neutrino Oscillation Program in the Fermilab Booster Neutrino Beam}",
    eprint = "1503.01520",
    archivePrefix = "arXiv",
    primaryClass = "physics.ins-det",
    reportNumber = "FERMILAB-PUB-15-0726-AD-PPD",
    month = "3",
    year = "2015"
}

@article{DUNE:2021tad,
    author = "Hewes, V. and others",
    collaboration = "DUNE",
    title = "{Deep Underground Neutrino Experiment (DUNE) Near Detector Conceptual Design Report}",
    eprint = "2103.13910",
    archivePrefix = "arXiv",
    primaryClass = "physics.ins-det",
    reportNumber = "FERMILAB-PUB-21-067-E-LBNF-PPD-SCD-T",
    doi = "10.3390/instruments5040031",
    journal = "Instruments",
    volume = "5",
    number = "4",
    pages = "31",
    year = "2021"
}

@inproceedings{Apyan:2022tsd,
    author = "Apyan, Aram and others",
    title = "{DarkQuest: A dark sector upgrade to SpinQuest at the 120 GeV Fermilab Main Injector}",
    booktitle = "{Snowmass 2021}",
    eprint = "2203.08322",
    archivePrefix = "arXiv",
    primaryClass = "hep-ex",
    reportNumber = "FERMILAB-CONF-22-175-PPD-SCD-T",
    month = "3",
    year = "2022"
}

@article{SHiP:2021nfo,
    author = "Ahdida, C. and others",
    collaboration = "SHiP",
    title = "{The SHiP experiment at the proposed CERN SPS Beam Dump Facility}",
    eprint = "2112.01487",
    archivePrefix = "arXiv",
    primaryClass = "physics.ins-det",
    doi = "10.1140/epjc/s10052-022-10346-5",
    journal = "Eur. Phys. J. C",
    volume = "82",
    number = "5",
    pages = "486",
    year = "2022"
}

@article{Capozzi:2024pmh,
    author = "Capozzi, Francesco and Dutta, Bhaskar and Gurung, Gajendra and Jang, Wooyoung and Shoemaker, Ian M. and Thompson, Adrian and Yu, Jaehoon",
    title = "{Enhancing the sensitivity to seesaw mechanism predictions in gauged B-L scenarios}",
    eprint = "2410.08981",
    archivePrefix = "arXiv",
    primaryClass = "hep-ph",
    reportNumber = "MI-HET-752",
    doi = "10.1103/PhysRevD.111.055036",
    journal = "Phys. Rev. D",
    volume = "111",
    number = "5",
    pages = "055036",
    year = "2025"
}

@article{Berryman:2019dme,
    author = "Berryman, Jeffrey M. and de Gouvea, Andre and Fox, Patrick J and Kayser, Boris Jules and Kelly, Kevin James and Raaf, Jennifer Lynne",
    title = "{Searches for Decays of New Particles in the DUNE Multi-Purpose Near Detector}",
    eprint = "1912.07622",
    archivePrefix = "arXiv",
    primaryClass = "hep-ph",
    reportNumber = "FERMILAB-PUB-19-607-ND-T, NUHEP-TH/19-16",
    doi = "10.1007/JHEP02(2020)174",
    journal = "JHEP",
    volume = "02",
    pages = "174",
    year = "2020"
}

@article{Orloff:2002de,
    author = "Orloff, J. and Rozanov, Alexandre N. and Santoni, C.",
    title = "{Limits on the mixing of tau neutrino to heavy neutrinos}",
    eprint = "hep-ph/0208075",
    archivePrefix = "arXiv",
    reportNumber = "PCCF-RI-0205",
    doi = "10.1016/S0370-2693(02)02769-7",
    journal = "Phys. Lett. B",
    volume = "550",
    pages = "8--15",
    year = "2002"
}

@article{Boiarska:2021yho,
    author = "Boiarska, Iryna and Boyarsky, Alexey and Mikulenko, Oleksii and Ovchynnikov, Maksym",
    title = "{Constraints from the CHARM experiment on heavy neutral leptons with tau mixing}",
    eprint = "2107.14685",
    archivePrefix = "arXiv",
    primaryClass = "hep-ph",
    doi = "10.1103/PhysRevD.104.095019",
    journal = "Phys. Rev. D",
    volume = "104",
    number = "9",
    pages = "095019",
    year = "2021"
}

@article{Barouki:2022bkt,
    author = "Barouki, Ryan and Marocco, Giacomo and Sarkar, Subir",
    title = "{Blast from the past II: Constraints on heavy neutral leptons from the BEBC WA66 beam dump experiment}",
    eprint = "2208.00416",
    archivePrefix = "arXiv",
    primaryClass = "hep-ph",
    doi = "10.21468/SciPostPhys.13.5.118",
    journal = "SciPost Phys.",
    volume = "13",
    pages = "118",
    year = "2022"
}

@article{ArgoNeuT:2021clc,
    author = "Acciarri, R. and others",
    collaboration = "ArgoNeuT",
    title = "{New Constraints on Tau-Coupled Heavy Neutral Leptons with Masses mN=280{\textendash}970{\,}{\,}MeV}",
    eprint = "2106.13684",
    archivePrefix = "arXiv",
    primaryClass = "hep-ex",
    reportNumber = "FERMILAB-PUB-21-296-ND-T",
    doi = "10.1103/PhysRevLett.127.121801",
    journal = "Phys. Rev. Lett.",
    volume = "127",
    number = "12",
    pages = "121801",
    year = "2021"
}

@article{ICARUS:2023gpo,
    author = "Abratenko, P. and others",
    collaboration = "ICARUS",
    title = "{ICARUS at the Fermilab Short-Baseline Neutrino program: initial operation}",
    eprint = "2301.08634",
    archivePrefix = "arXiv",
    primaryClass = "hep-ex",
    reportNumber = "FERMILAB-PUB-23-010-ND",
    doi = "10.1140/epjc/s10052-023-11610-y",
    journal = "Eur. Phys. J. C",
    volume = "83",
    number = "6",
    pages = "467",
    year = "2023"
}

@inproceedings{Batell:2022xau,
    author = "Batell, Brian and others",
    title = "{Dark Sector Studies with Neutrino Beams}",
    booktitle = "{Snowmass 2021}",
    eprint = "2207.06898",
    archivePrefix = "arXiv",
    primaryClass = "hep-ph",
    reportNumber = "FERMILAB-FN-1180-ND-T",
    doi = "10.2172/1882578",
    month = "7",
    year = "2022"
}

@article{Minkowski:1977sc,
    author = "Minkowski, Peter",
    title = "{$\mu \to e\gamma$ at a Rate of One Out of $10^{9}$ Muon Decays?}",
    reportNumber = "Print-77-0182 (BERN)",
    doi = "10.1016/0370-2693(77)90435-X",
    journal = "Phys. Lett. B",
    volume = "67",
    pages = "421--428",
    year = "1977"
}

@article{Mohapatra:1979ia,
    author = "Mohapatra, Rabindra N. and Senjanovic, Goran",
    title = "{Neutrino Mass and Spontaneous Parity Nonconservation}",
    reportNumber = "MDDP-TR-80-060, MDDP-PP-80-105, CCNY-HEP-79-10",
    doi = "10.1103/PhysRevLett.44.912",
    journal = "Phys. Rev. Lett.",
    volume = "44",
    pages = "912",
    year = "1980"
}

@article{Yanagida:1979as,
    author = "Yanagida, Tsutomu",
    editor = "Sawada, Osamu and Sugamoto, Akio",
    title = "{Horizontal gauge symmetry and masses of neutrinos}",
    reportNumber = "KEK-79-18-95",
    journal = "Conf. Proc. C",
    volume = "7902131",
    pages = "95--99",
    year = "1979"
}

@article{Gell-Mann:1979vob,
    author = "Gell-Mann, Murray and Ramond, Pierre and Slansky, Richard",
    title = "{Complex Spinors and Unified Theories}",
    eprint = "1306.4669",
    archivePrefix = "arXiv",
    primaryClass = "hep-th",
    reportNumber = "PRINT-80-0576",
    journal = "Conf. Proc. C",
    volume = "790927",
    pages = "315--321",
    year = "1979"
}

@article{Berlin:2018pwi,
    author = "Berlin, Asher and Gori, Stefania and Schuster, Philip and Toro, Natalia",
    title = "{Dark Sectors at the Fermilab SeaQuest Experiment}",
    eprint = "1804.00661",
    archivePrefix = "arXiv",
    primaryClass = "hep-ph",
    reportNumber = "SLAC-PUB-17238",
    doi = "10.1103/PhysRevD.98.035011",
    journal = "Phys. Rev. D",
    volume = "98",
    number = "3",
    pages = "035011",
    year = "2018"
}

@article{Batell:2020vqn,
    author = "Batell, Brian and Evans, Jared A. and Gori, Stefania and Rai, Mudit",
    title = "{Dark Scalars and Heavy Neutral Leptons at DarkQuest}",
    eprint = "2008.08108",
    archivePrefix = "arXiv",
    primaryClass = "hep-ph",
    doi = "10.1007/JHEP05(2021)049",
    journal = "JHEP",
    volume = "05",
    pages = "049",
    year = "2021"
}

@misc{bolton,
  author = {Bolton, Patrick D.},
  title  = "{Sterile Neutrino/Heavy Neutral Lepton Constraints}",
  howpublished = {\url{https://www.hep.ucl.ac.uk/~pbolton/}}
}

@misc{hostert,
  author = {Hostert, Matheus},
  title  = "{Heavy-Neutrino-Limits}",
  howpublished = {\url{https://github.com/mhostert/Heavy-Neutrino-Limits}}
}

@article{Alloul:2013bka,
    author = "Alloul, Adam and Christensen, Neil D. and Degrande, C{\'e}line and Duhr, Claude and Fuks, Benjamin",
    title = "{FeynRules  2.0 - A complete toolbox for tree-level phenomenology}",
    eprint = "1310.1921",
    archivePrefix = "arXiv",
    primaryClass = "hep-ph",
    reportNumber = "CERN-PH-TH-2013-239, MCNET-13-14, IPPP-13-71, DCPT-13-142, PITT-PACC-1308",
    doi = "10.1016/j.cpc.2014.04.012",
    journal = "Comput. Phys. Commun.",
    volume = "185",
    pages = "2250--2300",
    year = "2014"
}

@article{Esteban:2024eli,
    author = "Esteban, Ivan and Gonzalez-Garcia, M. C. and Maltoni, Michele and Martinez-Soler, Ivan and Pinheiro, Jo{\~a}o Paulo and Schwetz, Thomas",
    title = "{NuFit-6.0: updated global analysis of three-flavor neutrino oscillations}",
    eprint = "2410.05380",
    archivePrefix = "arXiv",
    primaryClass = "hep-ph",
    reportNumber = "IFT-UAM/CSIC-24-140, YITP-SB-2024-24, IPPP/24/64, IPPP/24/64, IFT-UAM/CSIC-24-140, YITP-SB-2024-24",
    doi = "10.1007/JHEP12(2024)216",
    journal = "JHEP",
    volume = "12",
    pages = "216",
    year = "2024"
}

@article{deNiverville:2016rqh,
    author = "deNiverville, Patrick and Chen, Chien-Yi and Pospelov, Maxim and Ritz, Adam",
    title = "{Light dark matter in neutrino beams: production modelling and scattering signatures at MiniBooNE, T2K and SHiP}",
    eprint = "1609.01770",
    archivePrefix = "arXiv",
    primaryClass = "hep-ph",
    doi = "10.1103/PhysRevD.95.035006",
    journal = "Phys. Rev. D",
    volume = "95",
    number = "3",
    pages = "035006",
    year = "2017"
}

@article{GEANT4:2002zbu,
    author = "Agostinelli, S. and others",
    collaboration = "GEANT4",
    title = "{GEANT4--a simulation toolkit}",
    reportNumber = "SLAC-PUB-9350, FERMILAB-PUB-03-339, CERN-IT-2002-003",
    doi = "10.1016/S0168-9002(03)01368-8",
    journal = "Nucl. Instrum. Meth. A",
    volume = "506",
    pages = "250--303",
    year = "2003"
}

@article{Drewes:2022akb,
    author = "Drewes, Marco and Klari{\'c}, Juraj and L{\'o}pez-Pav{\'o}n, Jacobo",
    title = "{New benchmark models for heavy neutral lepton searches}",
    eprint = "2207.02742",
    archivePrefix = "arXiv",
    primaryClass = "hep-ph",
    doi = "10.1140/epjc/s10052-022-11100-7",
    journal = "Eur. Phys. J. C",
    volume = "82",
    number = "12",
    pages = "1176",
    year = "2022"
}

@article{Bauer:2018onh,
    author = "Bauer, Martin and Foldenauer, Patrick and Jaeckel, Joerg",
    title = "{Hunting All the Hidden Photons}",
    eprint = "1803.05466",
    archivePrefix = "arXiv",
    primaryClass = "hep-ph",
    doi = "10.1007/JHEP07(2018)094",
    journal = "JHEP",
    volume = "07",
    pages = "094",
    year = "2018"
}

@article{Deppisch:2015qwa,
    author = "Deppisch, Frank F. and Dev, P. S. Bhupal and Pilaftsis, Apostolos",
    title = "{Neutrinos and Collider Physics}",
    eprint = "1502.06541",
    archivePrefix = "arXiv",
    primaryClass = "hep-ph",
    reportNumber = "MAN-HEP-2014-15",
    doi = "10.1088/1367-2630/17/7/075019",
    journal = "New J. Phys.",
    volume = "17",
    number = "7",
    pages = "075019",
    year = "2015"
}

@article{Cai:2017mow,
    author = "Cai, Yi and Han, Tao and Li, Tong and Ruiz, Richard",
    title = "{Lepton Number Violation: Seesaw Models and Their Collider Tests}",
    eprint = "1711.02180",
    archivePrefix = "arXiv",
    primaryClass = "hep-ph",
    reportNumber = "PITT-PACC-1712, IPPP-17-74, COEPP-MN-17-17",
    doi = "10.3389/fphy.2018.00040",
    journal = "Front. in Phys.",
    volume = "6",
    pages = "40",
    year = "2018"
}

@article{Bolton:2019pcu,
    author = "Bolton, Patrick D. and Deppisch, Frank F. and Dev, P. S. Bhupal",
    title = "{Neutrinoless double beta decay versus other probes of heavy sterile neutrinos}",
    eprint = "1912.03058",
    archivePrefix = "arXiv",
    primaryClass = "hep-ph",
    doi = "10.1007/JHEP03(2020)170",
    journal = "JHEP",
    volume = "03",
    pages = "170",
    year = "2020"
}

@article{deGouvea:2015euy,
    author = "de Gouv{\^e}a, Andr{\'e} and Kobach, Andrew",
    title = "{Global Constraints on a Heavy Neutrino}",
    eprint = "1511.00683",
    archivePrefix = "arXiv",
    primaryClass = "hep-ph",
    reportNumber = "NUHEP-TH-15-03",
    doi = "10.1103/PhysRevD.93.033005",
    journal = "Phys. Rev. D",
    volume = "93",
    number = "3",
    pages = "033005",
    year = "2016"
}

@article{Amrith:2018yfb,
    author = "Amrith, S. and Butterworth, J. M. and Deppisch, F. F. and Liu, W. and Varma, A. and Yallup, D.",
    title = "{LHC Constraints on a $B-L$ Gauge Model using Contur}",
    eprint = "1811.11452",
    archivePrefix = "arXiv",
    primaryClass = "hep-ph",
    reportNumber = "MCnet-18-30",
    doi = "10.1007/JHEP05(2019)154",
    journal = "JHEP",
    volume = "05",
    pages = "154",
    year = "2019"
}

@article{Deppisch:2018eth,
    author = "Deppisch, Frank F. and Liu, Wei and Mitra, Manimala",
    title = "{Long-lived Heavy Neutrinos from Higgs Decays}",
    eprint = "1804.04075",
    archivePrefix = "arXiv",
    primaryClass = "hep-ph",
    reportNumber = "IP/BBSR/2018-4, IP-BBSR-2018-4",
    doi = "10.1007/JHEP08(2018)181",
    journal = "JHEP",
    volume = "08",
    pages = "181",
    year = "2018"
}

@article{Basso:2008iv,
    author = "Basso, Lorenzo and Belyaev, Alexander and Moretti, Stefano and Shepherd-Themistocleous, Claire H.",
    title = "{Phenomenology of the minimal B-L extension of the Standard model: Z' and neutrinos}",
    eprint = "0812.4313",
    archivePrefix = "arXiv",
    primaryClass = "hep-ph",
    reportNumber = "SHEP-08-13",
    doi = "10.1103/PhysRevD.80.055030",
    journal = "Phys. Rev. D",
    volume = "80",
    pages = "055030",
    year = "2009"
}

@article{Chrzaszcz:2019inj,
    author = "Chrzaszcz, Marcin and Drewes, Marco and Gonzalo, Tom{\'a}s E. and Harz, Julia and Krishnamurthy, Suraj and Weniger, Christoph",
    title = "{A frequentist analysis of three right-handed neutrinos with GAMBIT}",
    eprint = "1908.02302",
    archivePrefix = "arXiv",
    primaryClass = "hep-ph",
    reportNumber = "gambit-physics-2019",
    doi = "10.1140/epjc/s10052-020-8073-9",
    journal = "Eur. Phys. J. C",
    volume = "80",
    number = "6",
    pages = "569",
    year = "2020"
}

@article{WA66:1985mfx,
    author = "Cooper-Sarkar, Amanda M. and others",
    collaboration = "WA66",
    title = "{Search for Heavy Neutrino Decays in the {BEBC} Beam Dump Experiment}",
    reportNumber = "CERN-EP-85-104",
    doi = "10.1016/0370-2693(85)91493-5",
    journal = "Phys. Lett. B",
    volume = "160",
    pages = "207--211",
    year = "1985"
}

@article{BaBar:2022cqj,
    author = "Lees, J. P. and others",
    collaboration = "BaBar",
    title = "{Search for heavy neutral leptons using tau lepton decays at BaBaR}",
    eprint = "2207.09575",
    archivePrefix = "arXiv",
    primaryClass = "hep-ex",
    reportNumber = "BABAR-PUB-22/002, SLAC-PUB-17695",
    doi = "10.1103/PhysRevD.107.052009",
    journal = "Phys. Rev. D",
    volume = "107",
    number = "5",
    pages = "052009",
    year = "2023"
}

@article{Feng:2024zfe,
    author = "Feng, Jonathan L. and Hewitt, Alec and Kling, Felix and La Rocco, Daniel",
    title = "{Simulating heavy neutral leptons with general couplings at collider and fixed target experiments}",
    eprint = "2405.07330",
    archivePrefix = "arXiv",
    primaryClass = "hep-ph",
    doi = "10.1103/PhysRevD.110.035029",
    journal = "Phys. Rev. D",
    volume = "110",
    number = "3",
    pages = "035029",
    year = "2024"
}

@article{Adhikary:2024nlv,
    author = "Adhikary, Jyotismita and others",
    title = "{Scientific program for the Forward Physics Facility}",
    eprint = "2411.04175",
    archivePrefix = "arXiv",
    primaryClass = "hep-ex",
    reportNumber = "DESY-24-174",
    doi = "10.1140/epjc/s10052-025-14048-6",
    journal = "Eur. Phys. J. C",
    volume = "85",
    number = "4",
    pages = "430",
    year = "2025"
}

@article{Gorbunov:2014wqa,
    author = "Gorbunov, D. and Makarov, A. and Timiryasov, I.",
    title = "{Decaying light particles in the SHiP experiment: Signal rate estimates for hidden photons}",
    eprint = "1411.4007",
    archivePrefix = "arXiv",
    primaryClass = "hep-ph",
    reportNumber = "INR-TH-2014-026, INR-TH/2014-026",
    doi = "10.1103/PhysRevD.91.035027",
    journal = "Phys. Rev. D",
    volume = "91",
    number = "3",
    pages = "035027",
    year = "2015"
}

@article{Foroughi-Abari:2021zbm,
    author = "Foroughi-Abari, Saeid and Ritz, Adam",
    title = "{Dark sector production via proton bremsstrahlung}",
    eprint = "2108.05900",
    archivePrefix = "arXiv",
    primaryClass = "hep-ph",
    doi = "10.1103/PhysRevD.105.095045",
    journal = "Phys. Rev. D",
    volume = "105",
    number = "9",
    pages = "095045",
    year = "2022"
}

@article{Kling:2025udr,
    author = "Kling, Felix and Reimitz, Peter and Ritz, Adam",
    title = "{Dark vector boson bremsstrahlung: New form factors for a broader class of models}",
    eprint = "2509.09437",
    archivePrefix = "arXiv",
    primaryClass = "hep-ph",
    reportNumber = "UCI-TR-2025-16",
    doi = "10.1103/fxpl-yhc2",
    journal = "Phys. Rev. D",
    volume = "113",
    number = "1",
    pages = "015008",
    year = "2026"
}

@article{Foroughi-Abari:2024xlj,
    author = "Foroughi-Abari, Saeid and Reimitz, Peter and Ritz, Adam",
    title = "{Closer look at dark vector splitting functions in proton bremsstrahlung}",
    eprint = "2409.09123",
    archivePrefix = "arXiv",
    primaryClass = "hep-ph",
    doi = "10.1103/fzlm-gsd7",
    journal = "Phys. Rev. D",
    volume = "112",
    number = "1",
    pages = "015030",
    year = "2025"
}

@article{Ilten:2018crw,
    author = "Ilten, Philip and Soreq, Yotam and Williams, Mike and Xue, Wei",
    title = "{Serendipity in dark photon searches}",
    eprint = "1801.04847",
    archivePrefix = "arXiv",
    primaryClass = "hep-ph",
    reportNumber = "MIT-CTP/4976, CERN-TH-2017-282, MIT-CTP-4976",
    doi = "10.1007/JHEP06(2018)004",
    journal = "JHEP",
    volume = "06",
    pages = "004",
    year = "2018"
}

@article{Dev:2023zts,
    author = "Dev, P. S. Bhupal and Dutta, Bhaskar and Han, Tao and Karthikeyan, Aparajitha and Kim, Doojin and Kim, Hyunyong",
    title = "{New physics at a neutron beam dump}",
    eprint = "2311.10078",
    archivePrefix = "arXiv",
    primaryClass = "hep-ph",
    reportNumber = "MI-HET-821, PITT-PACC-2320",
    doi = "10.1103/PhysRevD.110.L051703",
    journal = "Phys. Rev. D",
    volume = "110",
    number = "5",
    pages = "L051703",
    year = "2024"
}

@article{Kling:2018wct,
    author = "Kling, Felix and Trojanowski, Sebastian",
    title = "{Heavy Neutral Leptons at FASER}",
    eprint = "1801.08947",
    archivePrefix = "arXiv",
    primaryClass = "hep-ph",
    reportNumber = "UCI-TR-2017-18",
    doi = "10.1103/PhysRevD.97.095016",
    journal = "Phys. Rev. D",
    volume = "97",
    number = "9",
    pages = "095016",
    year = "2018"
}

\end{document}